\documentclass[journal]{IEEEtran}
\usepackage{array}
\usepackage{dirtytalk}
\usepackage{amsfonts} 
\newcolumntype{P}[1]{>{\centering\arraybackslash}p{#1}}
\usepackage{wrapfig}
\usepackage{cite}
\usepackage{caption}
\usepackage{multirow}
\usepackage{subcaption}
\usepackage{booktabs,tabularx,graphicx}
\usepackage{amsmath}
\usepackage{stackengine}

\usepackage{amsmath}

\usepackage{algorithmic}

\usepackage{array}

\begin{document}

\title{Discrimination of Internal Faults and Other Transients in an Interconnected System with Power Transformers and Phase Angle Regulators}

\author{Pallav~Kumar~Bera,~\IEEEmembership{}
        Can~Isik,~\IEEEmembership{Senior Member,~IEEE,}
        and~Vajendra~Kumar~\IEEEmembership{}

\thanks{Manuscript received February 20, 2020; revised May 23, 2020; accepted July 10, 2020.\textit{(Corresponding author: Pallav Kumar Bera.)}}
\thanks{Pallav Kumar Bera \& Can Isik are with the Department
of Electrical Engineering and Computer Science, Syracuse University,
NY, USA (e-mail: pkbera@syr.edu; cisik@syr.edu).}
\thanks{Vajendra Kumar was with Indian Institute of Technology, Roorkee, India (email: kumarvajendra@gmail.com).}
\thanks {© 2020 IEEE.  Personal use of this material is permitted.  Permission from IEEE must be obtained for all other uses, in any current or future media, including reprinting/republishing this material for advertising or promotional purposes, creating new collective works, for resale or redistribution to servers or lists, or reuse of any copyrighted component of this work in other works.
}}

\markboth{IEEE SYSTEMS JOURNAL}%
{Bera \MakeLowercase{\textit{et al.}}: Discrimination of Internal Faults and Other Transients in an Interconnected System with Phase Angle Regulators and Power Transformers}

\maketitle

\begin{abstract}
This study solves the problem of accurate detection of internal faults and classification of transients in a 5-bus interconnected system for Phase Angle Regulators (PARs) and Power Transformers (PTs). The analysis prevents mal-operation of differential relays in case of transients other than faults which include magnetizing inrush, sympathetic inrush, external faults with Current Transformer (CT) saturation, capacitor switching, non-linear load switching, and ferroresonance. A gradient boosting classifier (GBC) is used to distinguish the internal faults from the transient disturbances based on 1.5 cycles of 3-phase differential currents registered by a change detector. After the detection of an internal fault, GBCs are used to locate the faulty unit (PT, PAR series, or exciting unit) and identify the type of fault. In case a transient disturbance is detected, another GBC classifies them into the six disturbances. Five most relevant frequency and time domain features obtained using Information Gain are used to train and test the classifiers. The proposed algorithm distinguishes the internal faults from the other transients with a balanced accuracy ($\bar{\eta}$) of 99.95\%. The faulty transformer unit is located with $\bar{\eta}$ of 99.5\%  and the different transient disturbances are identified with $\bar{\eta}$ of 99.3\%. {Moreover, the reliability of the scheme is verified for different rating and connection of the transformers involved, CT saturation, and noise level in the signals.} These GBC classifiers can work together with a conventional differential relay and offer a supervisory control over its operation.  PSCAD/EMTDC software is used for simulation of the transients and to develop the two and three-winding transformer models for creating the internal faults including inter-turn and inter-winding faults.
\end{abstract}


\IEEEpeerreviewmaketitle
\section{Introduction}
\IEEEPARstart{P}{ower} Transformers are an integral part of an electrical grid and their protection is vital for reliable and stable operation of the power system. An important requirement of the protection system is the faithful discrimination of faults from other transients. Differential protection has been the primary protection scheme in transformers because of its inherent selectivity and sensitivity. Mal-operations due to magnetizing and sympathetic inrush, and CT saturation during external faults are the major problems associated with differential protection. Second-harmonic restraint method is extensively used to distinguish internal faults from magnetizing inrush since more second-harmonic component exists in inrush currents than in internal faults \cite{harmonic}. However, higher second-harmonics are generated during internal faults with CT saturation, presence of shunt capacitance, or because of the distributed capacitance of EHV lines \cite{tx3pro}. In addition, the second-harmonic content in inrush currents has reduced in modern transformers with soft core material \cite{modern_core}. Hence, several cases of mal-operation of conventional relays in distinguishing faults and inrush have been reported \cite{china}.
CT saturation during external faults may also cause false trips due to the inefficient setting of commonly used dual-slope biased differential relays \cite{ctsat1}.

\begin{figure}[htpb]
\centerline{\includegraphics[width=3.55 in, height= 2.38 in]{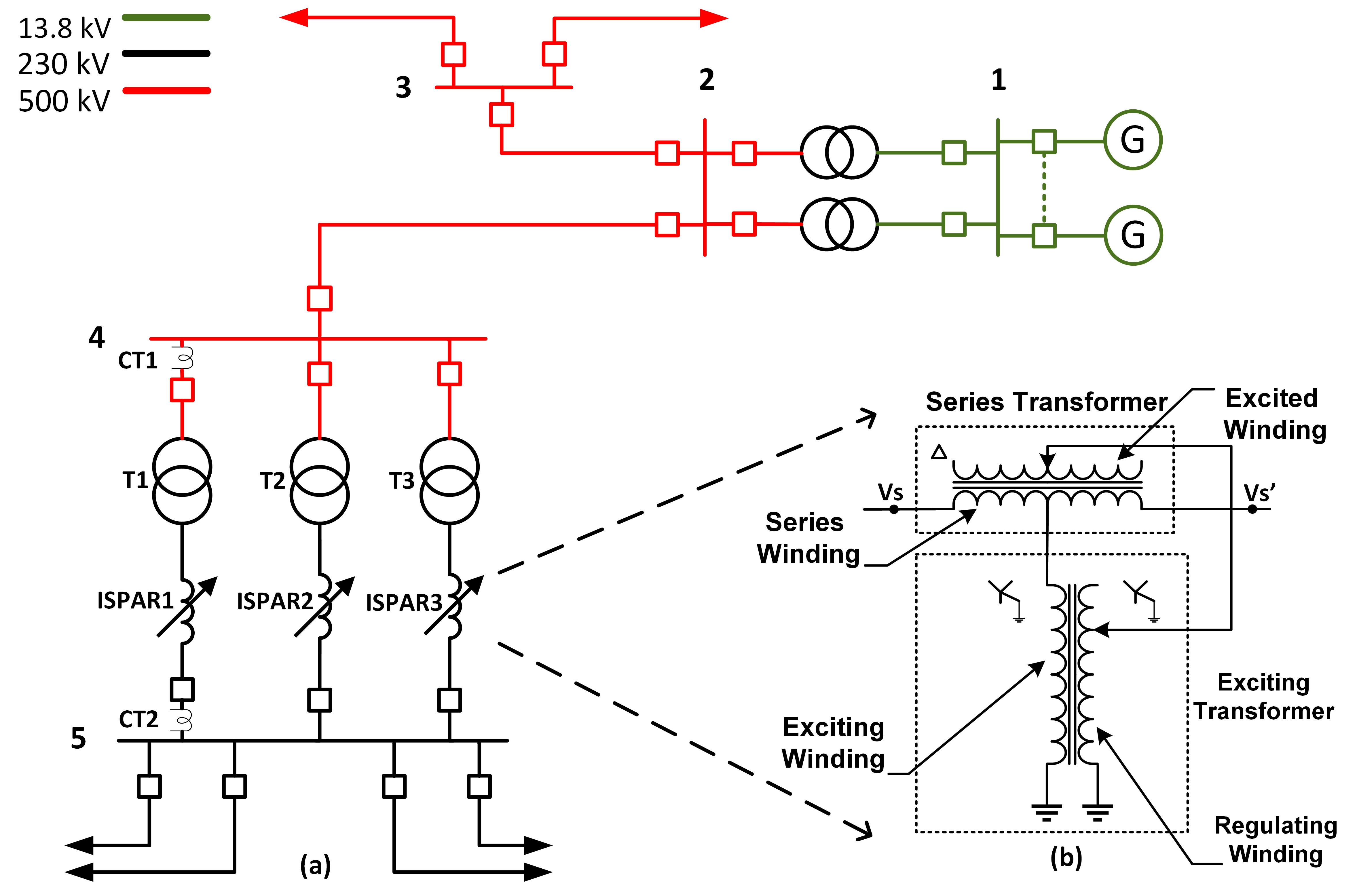}}
\vspace{0mm}\caption{(a) 5-bus interconnected system with ISPARs, PTs, T-lines, and AC sources, (b) Series and exciting transformers in ISPAR}
\label{leaps}
\end{figure}

Phase Angle Regulators or Phase Shifters or Phase Shift Transformers are a special class of transformers used to control real power flow in parallel transmission lines. They ensure the system reliability and allow easier integration of new generations with the grid. By regulating the phase angle between the sending and receiving ends they prevent overloading of a line and re-routes power via another line. PARs can be categorized on the basis of the number of cores and magnitude of sending end voltage with respect to the receiving end. Indirect Symmetrical Phase Angle Regulators (ISPAR) having the same sending and receiving end voltages with two transformer units, namely, series and exciting (Fig.\ref{leaps}(b)), has been chosen as one of the subjects (other being the PTs) in this study because of their popularity and security against higher voltage levels as the load tap changer (LTC) is not exposed to system disturbances. The exciting unit is responsible for creating the required phase difference to regulate the power which can be controlled by the LTC connections and an advance-retard-switch located on its secondary winding \cite{ibrahim}. The modified real power flow in a transmission line with a PAR is given by
\begin{equation}
    P=\frac {V_S \times V_L}{X_{line}+ X_{PAR}} \times sin(\theta + \alpha)
\end{equation} where, $V_S$ is source voltage, $V_L$ is load voltage; $\theta$ is the phase angle difference between $V_S$ and $V_L$;  $X_{line},X_{PAR}$ are the transmission line and PAR reactance respectively; and $\alpha$ is the new constraint added which is responsible for controlling the power flow.
The PARs similar to PTs require a fast, sensitive, secure, and dependable protection system. Discriminating external faults with CT saturation, magnetizing inrush, and other transient disturbances from internal faults is a challenge for the protection systems of PARs as well. Moreover, methods used to compensate the phase for differential relays in PTs with a fixed phase shift are not applicable in PARs with variable phase shift \cite{pstguide}.

Authors have used different intelligent methods to distinguish internal faults and magnetizing inrush in PTs in the past decade. A combination of Artificial Neural Network (ANN) and spectral energies of wavelet components was used to discriminate internal faults and inrush in \cite{ann1}. Support Vector Machines (SVM) and Decision Tree (DT) based transformer protection were proposed in \cite{SVM1,SVM2} and \cite{dt1,dt2,dtwt} respectively.
Probabilistic Neural Network (PNN) has been used to detect different conditions in PT operation in \cite{tripathyrb}. Random Forest Classifier (RFC) was proposed to discriminate internal faults and inrush in \cite{Shahrfc}. Works of literature also suggest extensive use of S-Transform, Hyperbolic S-Transform, Wavelet Transform (WT) to detect Power Quality (PQ) transient disturbances and then classify them using DT, SVM, ANN, PNN \cite{PQ1,PQ2,PQ3,PQ4,PQ5,PQ6}. These transient disturbances are caused by variations in load, capacitor switching, charging of transformers, starting of induction machines, use of non-linear loads, etc.
Contrarily, literature investigating internal faults and inrush in an ISPAR is limited. However, attempts were made in \cite{pallav} where internal faults are distinguished from magnetizing inrush using WT and then the internal faults are classified using ANN and in \cite{pallav2} where the internal faults in series and exciting transformers of the ISPAR are classified using RFC. Further, the authors have predominately used an isolated and simple network having a PT \cite{ann1,SVM1,SVM2,dt1,dt2,dtwt,tripathyrb} or a PAR \cite{pallav,pallav2} to support their proposed protection scheme. Also, the transient disturbances have not been studied rigorously in these works.

This paper studies the use of Decision Tree based algorithms to discriminate the internal faults and other transient disturbances including magnetizing inrush and CT saturation during external faults in a 5-bus interconnected system with Phase Angle Regulators and Power Transformers which has not been attempted before. Customized two-winding and three-winding transformers are developed to simulate the internal faults. A change detector has been used to detect and register the differential currents. Five most relevant time and frequency domain features have been used to train SVM, RFC, DT, and GBC classifiers to detect, locate and identify the internal faults and classify six transient disturbances. The proposed scheme is tested on 101,088 transients cases simulated on Power System Computer Aided Design (PSCAD)/ Electromagnetic Transients including DC (EMTDC) by varying various system parameters. The entire dataset of internal faults and other disturbances is made available on IEEE Dataport \cite{data}.

The rest of the paper is organized as follows. Section II illustrates the modeling and simulation of the internal faults and other transient disturbances in the power network containing PTs and ISPARs. Section III comprises the detection of internal faults, feature extraction and selection, and the classifiers used for the detection and identification of transients. Section IV includes the results for detection of internal faults, identification of faults and transient disturbances, and evaluates the effect of noise, CT saturation, and change in transformer rating and connection on the proposed scheme. Section V concludes the paper. 

\section{Modeling and Simulation}\label{sec2}
The power network chosen for the simulation of the internal faults and the transient disturbances is based on a proposed Pumped-storage (efficient form of renewable storage designed to meet energy needs and reduce emissions by utilizing the energy stored in an upper water body pumped from a lower water body) project in California, USA \cite{siemens}.

PSCAD/EMTDC is used for the modeling and simulation of the transients in the ISPAR and PT in the chosen interconnected power system. Fig.\ref{leaps}(a) shows the single-line diagram of the 5-bus interconnected model consisting of the AC source, transmission lines, ISPARs, PTs, and 3-phase loads working at 60Hz. The ISPARs have a rating of 500 MVA, 230kV/230kV, with phase angle variations of $\pm25^{\circ}$ and the PTs are rated at 500 MVA, 500kV/230kV. The AC source consists of 9 units of 120 MVA, 13.8kV hydro-generators. Two transformers are used in cascade to step up the voltage from 13.8kV to 500kV. 3 ISPARs (ISPAR$_1$, ISPAR$_2$, and  ISPAR$_3$) are connected between bus4 and bus5 through transformers T$_1$, T$_2$, and T$_3$. Only the internal faults in ISPAR$_1$ and T$_1$ are studied. 

The three-winding transformer required for the series units of ISPAR and the two-winding transformer required for the exciting units of ISPAR and PTs for the simulation of various internal faults including turn-to-turn and primary-to-secondary winding faults are developed in PSCAD/EMTDC with Fortran. The self-inductance terms ($Li$) and the mutual inductance terms ($Mij$) of the $4\times 4$ $L$ matrix (Eq.\ref{matrix}) of the single-phase two-winding transformer and $6\times6$ $L$ matrix of the single-phase three-winding transformer are evaluated from primary and secondary voltages, the magnetizing component of the no-load excitation current ($I_m$), and the short-circuit tests. The modeled components have the provision to change the saturation characteristics, \% of winding shorted and other parameters. The Fortran script of the two-winding transformer is shown in the Appendix.

\begin{equation}
\footnotesize
L=
\begin{bmatrix}
Lx & Mxy & Mxz& Mxw\\
Myx & Ly & Myz& Myw\\
Mzx & Mzy & Lz& Mzw\\
Mwx & Mwy & Mwz& Lw
\end{bmatrix}\label{matrix}
\end {equation}

The study covers various internal faults in the ISPAR and PT, capacitor switching, switching of non-linear loads, magnetizing inrush, sympathetic inrush, external faults with CT saturation, and ferroresonance. In the following paragraphs, these conditions are considered  one after the other. The simulation run-time, fault/disturbance inception time, and fault duration time are 15.2s, 15.0s, and 0.05s (3 cycles) respectively in all cases. The multi-run component is used to change the parameter values wherever possible to get the different simulation cases and snapshots of the first simulation runs are used to start the simulation from initialized conditions to reduce the simulation time. 
\subsection{Internal Faults}
The internal faults are created in the PT, ISPAR series, and ISPAR exciting unit. 88,128 internal fault cases which include basic internal faults, turn-to-turn, and winding-to-winding faults are simulated by varying the fault resistance, \% of winding shorted, fault inception time, forward or backward shift, and the LTC in the exciting unit.

\subsubsection{ Internal phase \& ground faults (ph \& g)}
Phase winding to ground ($w_{a}$-g, $w_{b}$-g, $w_{c}$-g), phase winding to phase winding to ground ($w_{a}$-$w_{b}$-g, $w_{a}$-$w_{c}$-g, $w_{b}$-$w_{c}$-g), phase winding to phase winding ($w_{a}$-$w_{b}$, $w_{a}$-$w_{c}$, $w_{b}$-$w_{c}$), 3-phase winding ($w_{a}$-$w_{b}$-$w_{c}$), and 3-phase winding to ground ($w_{a}$-$w_{b}$-$w_{c}$-g) faults are simulated in the primary (P) and secondary (S) sides of the PT and on the primary and secondary sides of exciting and series transformer units in the ISPAR. Table \ref{tabintrnalfault} shows the values of different system and fault parameters in T$_1$ and ISPAR$_1$ (Fig.\ref{leaps}(a)) which are varied to get the training and testing cases for the internal phase \& ground faults. 

\begin{table}[htbp]
    \centering
    \caption{Parameters for ph \& g  faults in the ISPAR and PT}
    \label{tabintrnalfault}
    \begin{tabular}{ll}
\hline
Variables  & Values \\ \hline
Fault resistance &  0.01, 0.5 \&   10 $\Omega$     (3)   \\
\% of winding shorted & 20\%, 50\%, 80\%  (3)\\
Fault type      & $w$-g, $w$-$w$-g, $w$-$w$, $w$-$w$-$w$ \& $w$-$w$-$w$-g (11) \\
Fault inception time       & 15s to 15.0153s in steps of 1.38ms (12)\\
Fault location            &     \begin{tabular}{@{\extracolsep{\fill}}l}Transformer (P \& S)  (2) \\ ISPAR Exciting unit (P \& S) (2)\\ \& ISPAR Series unit (P \& S) (2) \end{tabular}  \\
Phase shift & Forward and backward  (2)\\ 
LTC  & 0.2,0.4,0.6,0.8,1[1 \& 0.5 in ISPAR exciting] \\ \midrule
 \multicolumn{2}{l}{{}}                                                 \\
\multicolumn{2}{l}{\multirow{-2}{*}{{ \begin{tabular}[c]{@{}l@{}} \footnotesize Transformer or ISPAR series faults = 3$\times$3$\times$11$\times$12$\times$2$\times$2$\times$5 = 23,760                                                                                                                                                                                   \\ \footnotesize ISPAR exciting faults = 3$\times$3$\times$11$\times$12$\times$2$\times$2$\times$2 = 9504\end{tabular}}}} \\ \hline
\end{tabular}
\end{table}

\subsubsection{Turn-to-turn (T-T) faults} 
\begin{figure}[htbp]
\centerline{\includegraphics[width=3.2 in, height= 2.7 in]{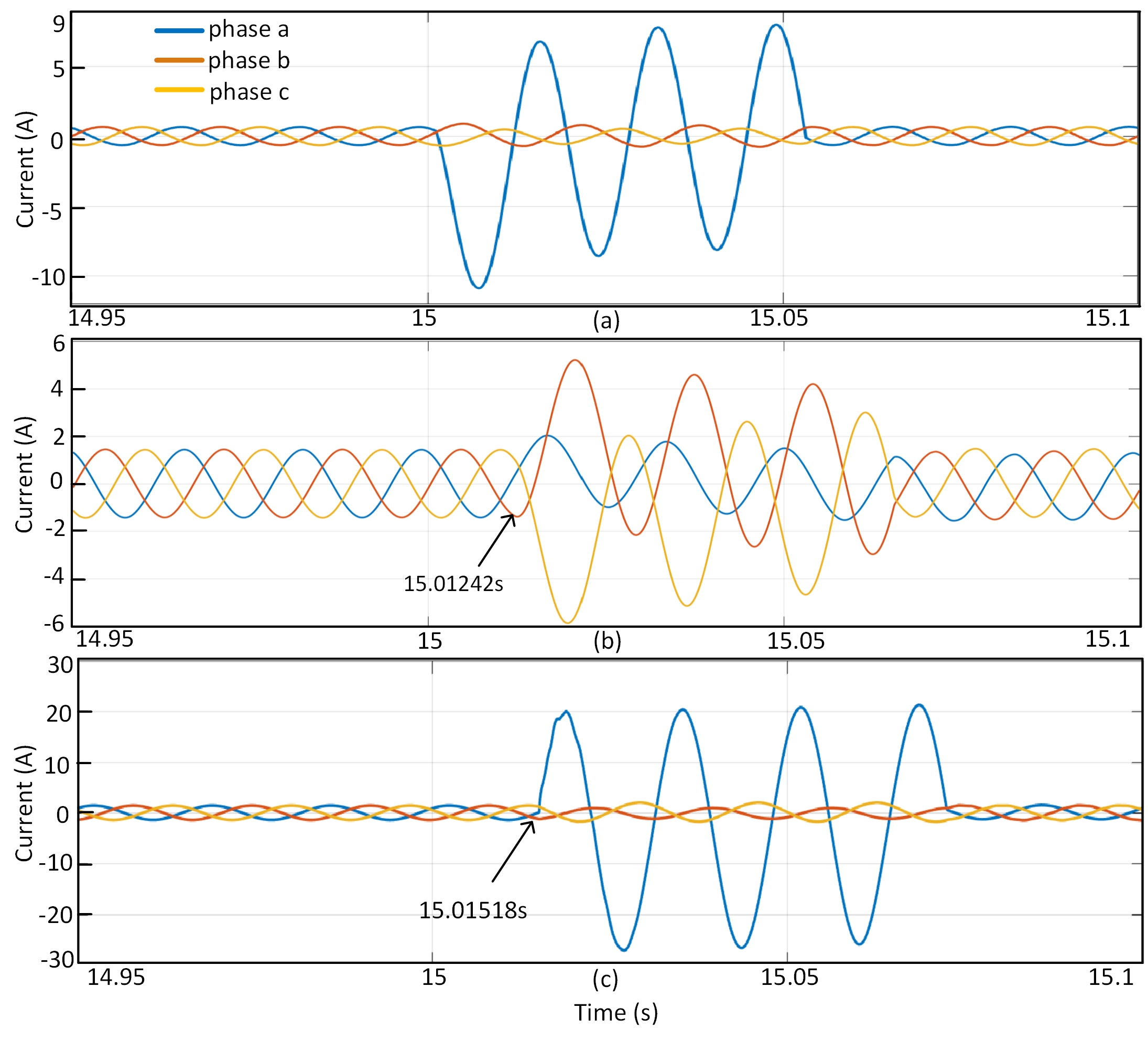}}
\vspace{0mm}\caption{3-phase differential currents for turn-to-turn faults in (a) primary of exciting unit, (b) primary of series unit, and (c) primary of PT}
\label{tt}
\end{figure}

About 70-80\% of faults in transformers are due to turn-to-turn insulation failures. Thermal, mechanical and electrical stress degrades the insulation and causes turn-to-turn faults which may lead to more serious faults and inter-winding faults if not detected quickly \cite{turn}. Table \ref{tab_ww_tt} shows the values of different parameters of the PT and the series and exciting unit of ISPAR used to simulate 20,736 turn-to-turn faults. Fig.\ref {tt}(a) shows the differential currents for LTC= half, fault inception time=15s, backward phase shift, fault resistance=0.01$\Omega$ and \% turns shorted=20 in primary of exciting unit. Fig.\ref {tt}(b) shows the differential currents for LTC= full, fault inception time=15.0124s, backward phase shift, fault resistance=0.01$\Omega$ and \% turns shorted=40 in primary of series unit. Fig.\ref {tt}(c) shows the differential currents for LTC= full, fault inception time=15.01518s, forward phase shift, fault resistance=0.01$\Omega$ and \% turns shorted=60 in primary of the PT.
\begin{table}[ht]
    \centering
    \caption{Parameters for winding-to-winding \& turn-to-turn faults in the ISPAR and PT}
    \label{tab_ww_tt}
    \begin{tabular}{ll}
\hline
Variables  & Values \\ \hline
Fault resistance &  0.01, 0.5 \&   10 $\Omega$     (3)   \\
\% of winding shorted & 20\%, 40\%, 60\%, 80\%  (4)\\
Fault inception time       & 15s to 15.0153s in steps of 1.38ms (12)\\
Fault location            &     \begin{tabular}{@{\extracolsep{\fill}}l}Transformer phase a,b,c (P \& S)  (6) \\ ISPAR Exciting phase a,b,c (P \& S) (6)\\ \& ISPAR Series phase a,b,c (P \& S) (6) \end{tabular}  \\
Phase shift & Forward and backward  (2)\\ 
LTC  & 0.2,0.4,0.6,0.8,1 [1 \& 0.5 in ISPAR exciting] \\ \midrule
\multicolumn{2}{l}{{ }}                                                                                                                                                                                                                                                             \\
\multicolumn{2}{l}{{ }}                                                                                                                                                                                                                                                                                  \\
\multicolumn{2}{l}{{ }}                                                                                                                                                                                                \\
\multicolumn{2}{l}{\multirow{-4}{*}{{ \begin{tabular}[c]{@{}l@{}}\footnotesize Transformer or ISPAR series(T-T) faults = 3$\times$4$\times$12$\times$6$\times$2$\times$5 = 8640\\  \footnotesize ISPAR exciting(T-T) faults = 3$\times$4$\times$12$\times$6$\times$2$\times$2 = 3456 \\  \footnotesize Transformer or ISPAR series(W-W) faults = 3$\times$4$\times$12$\times$3$\times$2$\times$5= 4320\\  \footnotesize  ISPAR exciting(W-W) faults = 3$\times$4$\times$12$\times$3$\times$2$\times$2 = 1728\end{tabular}}}} \\ \hline
\end{tabular}
\end{table}
\subsubsection{Winding-to-winding (W-W) faults}
The electrical, thermal and mechanical stress due to short circuits and transformer aging reduces the mechanical and dielectric strength of the winding and results in degradation of the insulation between LV and HV winding and may damage the winding eventually \cite{turn}. Table \ref{tab_ww_tt} shows the values of different parameters of the PT and the series and exciting unit of ISPAR used to simulate 10,368 winding-to-winding faults.

\subsection{Magnetizing inrush} 
\begin{figure}[htbp]
\centerline{\includegraphics[width=3.3 in, height= 1.8 in]{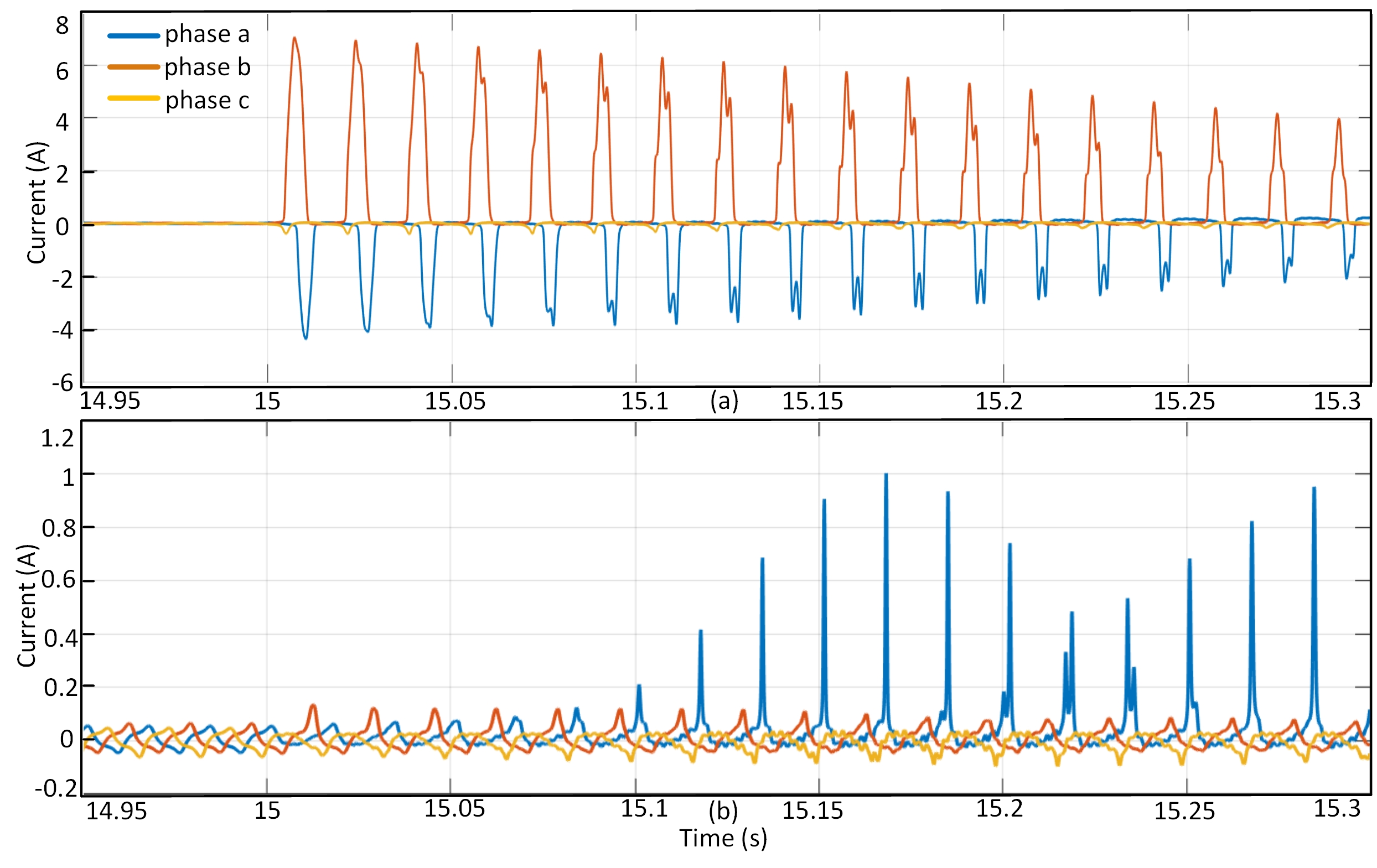}}
\vspace{0mm}\caption{ 3-phase differential currents for (a) Magnetizing inrush, and (b) Sympathetic inrush}
\label{inr}
\end{figure}

Transients caused by the energization of transformers are common and discrimination of inrush from fault currents has been studied since the 19th century. The harmonic restraint relays fail to detect inrush currents in transformers with modern core materials. The flux in a transformer core just after switching can be expressed as
\begin{equation}\phi = \phi_R + \phi_m cos\omega t' - \phi_m cos\omega( t + t')\end{equation}
where, $\phi_R$ {represents} residual flux,  $\phi_m$ {represents} maximum flux, and t$'$ {is} switching time. The transformer draws a high peaky non-sinusoidal
\begin{figure}[htbp]
\centerline{\includegraphics[width=2.0 in, height= 1.8 in]{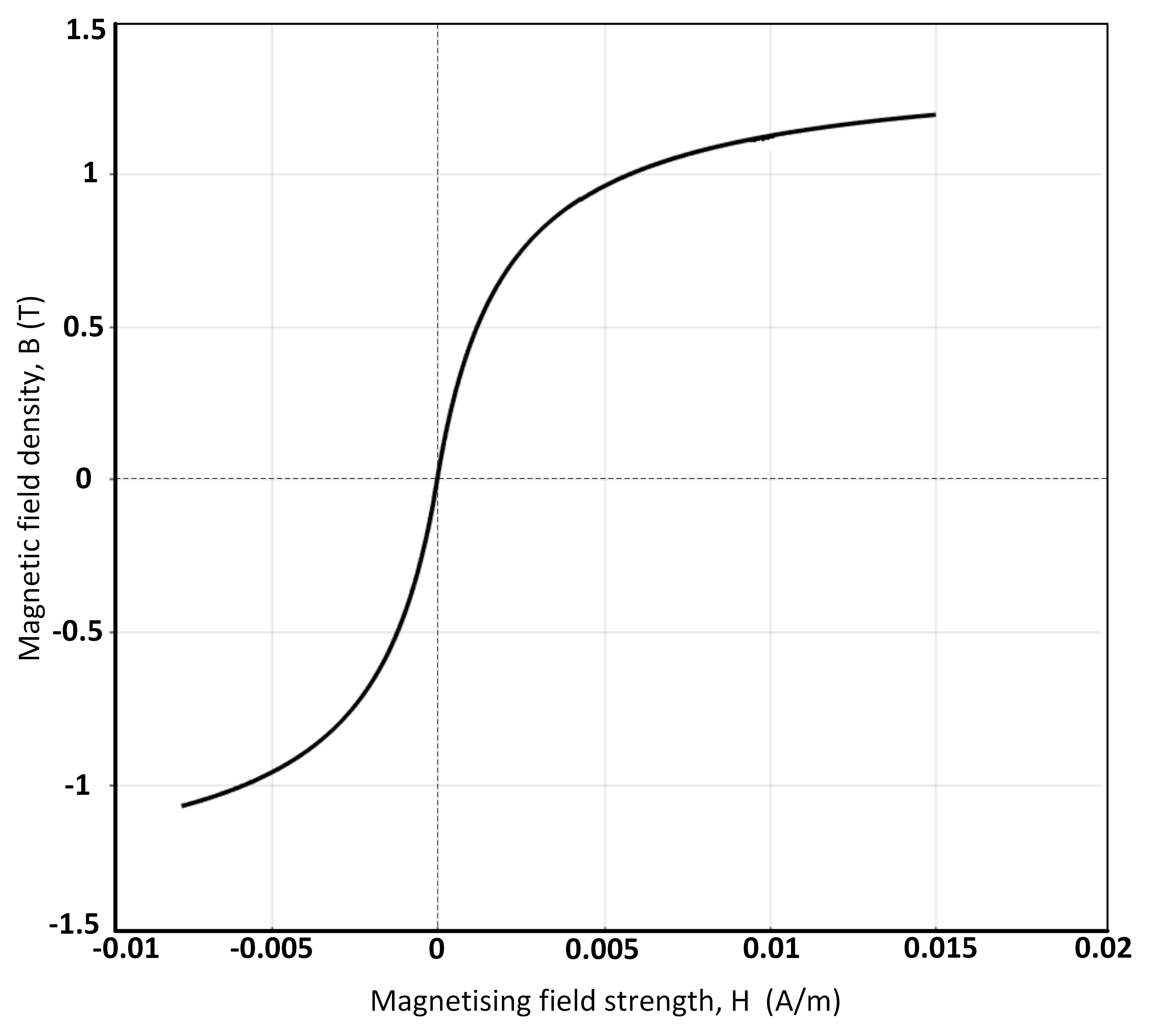}}\vspace{0mm}
\caption{\scriptsize{ B-H curve of transformer core}}
\label{bh}
\end{figure}
current to meet the high flux demand when switched on. Since this current flows only on one side of the transformer the differential scheme mal-operates. T$_1$ (Fig.\ref{leaps}(a)) is the incoming 3-phase transformer and DC sources are used to get the desired $\phi_R$ in the single-phase transformers. The values for the DC currents in phase-a, b, and c are obtained from the B-H curve of the transformer core material as shown in Fig.\ref{bh}. Table \ref{tab23} shows the values of different parameters including $\phi_R$ and  t$'$ used to get the data for training and testing for magnetizing inrush and Fig.\ref {inr}(a) shows the 3-phase differential currents for LTC = full, switching time = 15s, forward phase shift, and -80\% residual flux.
\begin{table}[htbp]
\centering
\caption{Parameters for Magnetizing and Sympathetic inrush\label{tab23}}
{\begin{tabular}{ll}\toprule
Variables  & Values \\ \midrule
Residual flux &  $\pm80\%, \pm40\%, 0\% $ in 3 phases; $5\times3$ = (15)\\
Switching time       & 15s to 15.0153s in steps of 1.38ms   (12)     \\
LTC  & 0.2 to full tap in steps of 0.2 (5)                       \\
Phase shift & Forward and backward (2)\\
&  {Total=$15\times12\times5\times2$=1800}\\
\hline
\end{tabular}}{}
\end{table}

\subsection{Sympathetic Inrush} Sympathetic inrush occurs in the in-service transformer (T$_1$) when the incoming transformer (T$_2$) is energized in a resistive network at no-load. The asymmetrical flux change per cycle during switching of T$_2$ which drives T$_1$ to saturation can be expressed as
\begin{equation}\Delta\phi=\int_{t}^{2\pi + t}[(R_{sys} + R_{T_1})i_1 + R_{sys} i_2]\end{equation}  
where $R_{sys}$ {is the} system resistance , and $R_{T_1}$ {is the} resistance of transformer T$_1$, $i_1$ and $i_2$ are magnetizing currents of T$_1$ and T$_2$.  This interaction between the incoming and the in-service transformers leads to mal-operation of differential relays of the in-service transformer due to failure of harmonic restraint relays and may cause prolonged harmonic over-voltages \cite{sym2}. The use of superconducting winding, soft magnetic material in the core, and CT local transient saturation are some factors responsible for these mal-operations \cite{modern_core} \cite{ctsat}. Sympathetic inrush is influenced by the residual flux ($\phi_R$) of the incoming transformer, switching time (t$'$), and system resistance \cite{kumbhar} and takes place with the incoming transformer energized in either series or parallel. The magnitude and direction of $\phi_R$, and t$'$ are altered and the incoming transformer T1 is connected in parallel to simulate the scenarios. Table \ref{tab23} shows the values of the different parameters used to get the training and testing cases for sympathetic inrush. Fig.\ref {inr}(b) shows the 3-phase differential currents for LTC = 0.2, switching time = 15s, forward phase shift, and -80\% residual flux.

\subsection{External faults with CT saturation}
\begin{figure}[htbp]
\centerline{\includegraphics[width=3.3 in, height= 2.8 in]{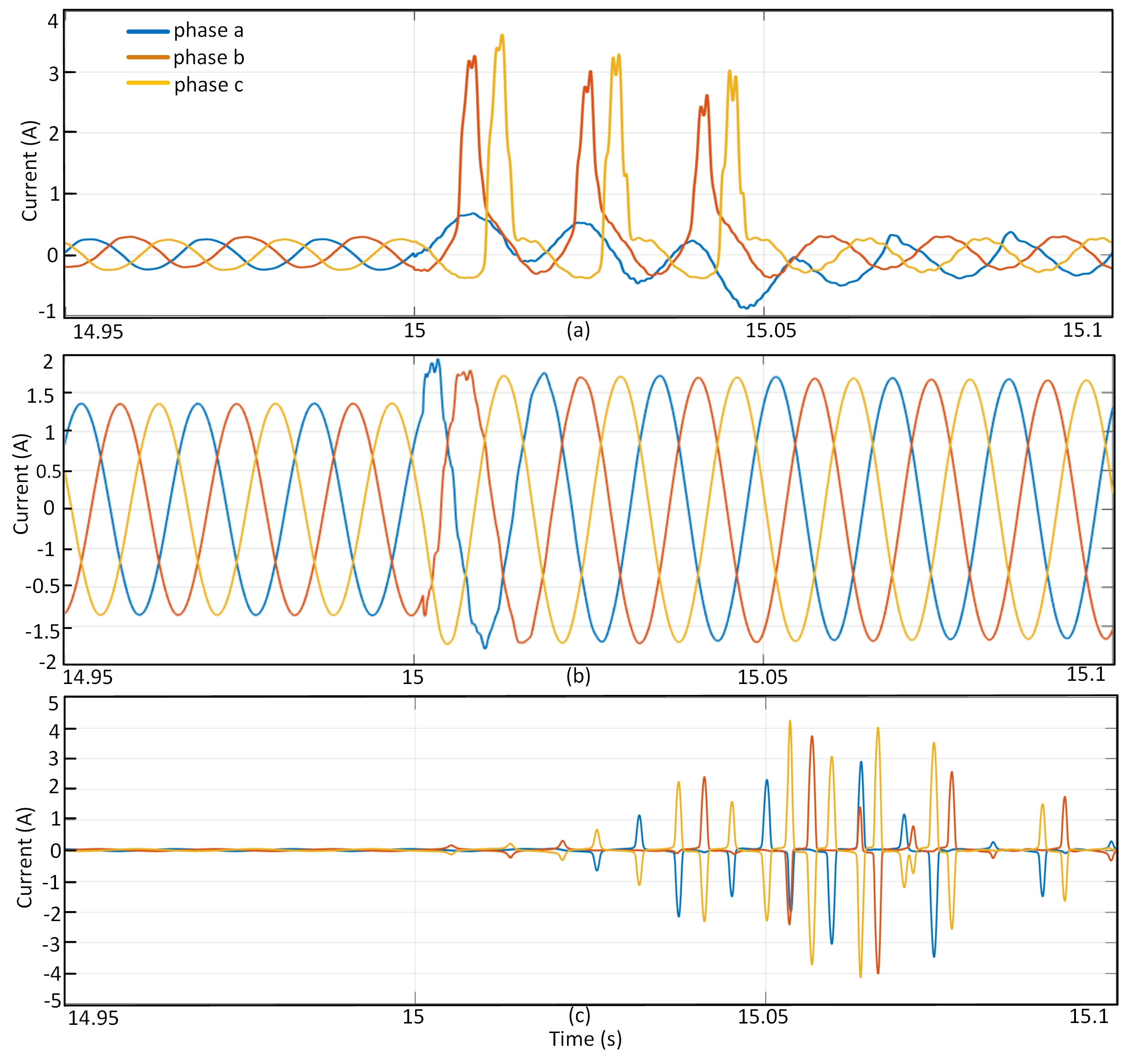}}
\vspace{0mm}\caption{3-phase differential currents for (a) External fault with CT saturation, (b) Capacitor Switching, and (c) Ferroresonance}
\label{ext}
\end{figure}

The differential currents become non-zero due to CT saturation in case of heavy through faults and may lead to a false trip. While raising the bias threshold ensures the security (i.e. no mal-operation), the dependability for in-zone resistive faults gets reduced. The external faults with CT saturation are simulated on the 500kV and 230kV buses (bus4 \& bus5). The values for the different parameters are given in Table \ref{externaltab}. Fig.\ref {ext}(a) shows the 3-phase differential currents for an external line-to-ground (lg) fault when LTC = 0.2,  phase shift = forward, fault inception time = 15s, and fault resistance = 0.01$\Omega$ on the 230kV bus.

\begin{table}[ht]
\centering
\caption{Parameters for External faults on 230kV \& 500kV bus}
\label{externaltab}
\begin{tabular}{ll}\toprule
Variables  & Values \\ \midrule
Fault resistance & 0.01, 0.5 \&   10 $\Omega$     (3)   \\
Fault type      & $l$g, $ll$g, $l$$l$, $l$$l$$l$ \& $lll$g (11) \\
Fault inception time       & 15s to 15.0153s in steps of 1.38ms    (12) \\
LTC  & 0.2 to full tap in steps of 0.2 (5)\\
Phase shift & Forward and backward  (2)\\ 
Fault location & 230kV \& 500kV bus (2)\\
& {Total=$3\times11\times12\times5\times2\times2$=7920}\\
\hline
\end{tabular}{}
\end{table}

\subsection{Non-linear Load Switching} 
\begin{figure}[htbp]
\centerline{\includegraphics[width=3.3 in, height=1.8 in]{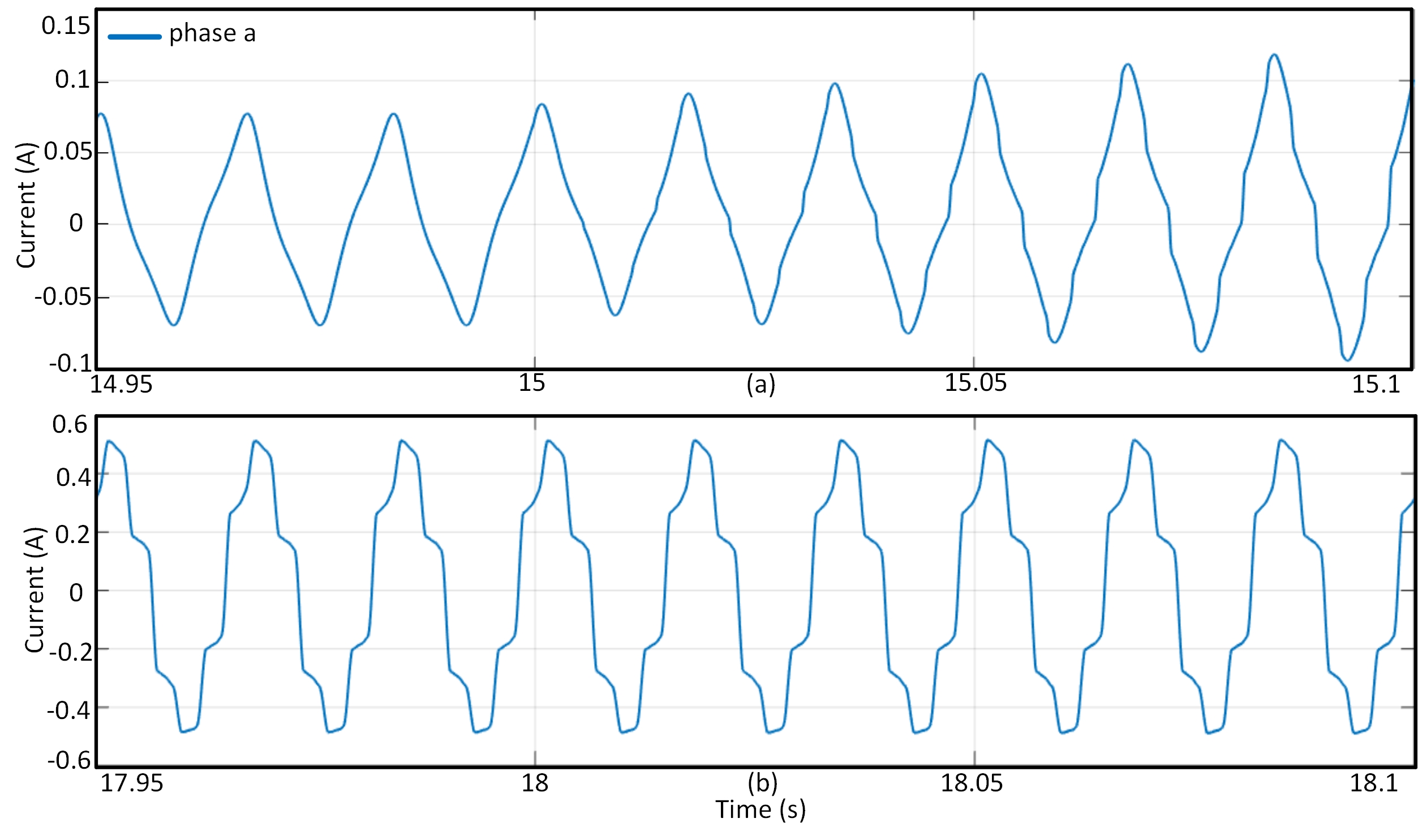}}
\vspace{0mm}\caption{Non-linear Load Switching (a) Transient, and (b) Steady-state differential currents}
\label{nl}
\end{figure}

With the advancement in semiconductor technology and the use of non-linear loads with power converters, harmonic contents in the line currents have increased. The differential relays may mal-operate when non-linear loads e.g steel furnaces are switched in a network containing transformers because of mutual enhancement effects between the transformer core and the load causing extreme saturation of the transformer core for several cycles \cite{non_linear_load_switch}. The harmonic information has been used to discriminate faults from other disturbances and locate the faults in the transmission line using SVM and ANN \cite{koley}. A thyristor-based 6-pulse bridge rectifier with a wye-delta transformer as the non-linear load is connected to the 230 kV bus to obtain the training and testing cases for load switching. The values for the different parameters are given in table \ref{tab33} and Fig.\ref {nl} shows the phase-a differential current for LTC = full, switching time = 15s and firing angle of 0$^{\circ}$. Fig.\ref {nl}(a) shows the transient and Fig.\ref {nl}(b) shows the steady-state differential current after the switching.

\begin{table}[ht]
\centering
\caption{Parameters for Non-linear Load Switching \label{tab33}}
{\begin{tabular}{ll}\toprule
Variables  & Values \\ \midrule
Firing angle & 0$^{\circ}$, 10$^{\circ}$, 20$^{\circ}$, 30$^{\circ}$, 40$^{\circ}$, 50$^{\circ}$ (6)                         \\
Switching time       & 15s to 15.0153s in steps of 1.38ms    (12)   \\
LTC  & 0.2 to full tap in steps of 0.2 (5)\\
& {Total = $6\times12\times5$ = 360}\\

\hline
\end{tabular}}{}
\end{table}

\subsection{Capacitor Switching} Capacitor banks are used to improve voltage profile, reduce losses, and enhance power factor. Mal-functioning of customer equipment due to voltage magnification coinciding with capacitor switching is common. \cite{5051} used WT to detect high transient inrush currents from capacitor-bank switching to avoid malfunctioning of instantaneous and time overcurrent relays (50/51). A capacitor bank having 3 Legs of 500 MVAr each is connected to the 230kV bus. Capacitor bank reactors and resistors are used in each Leg to reduce the effect of transients in voltages. Table \ref{tab143} shows the different parameters and their values used to get the data for training and testing for capacitor switching. Fig.\ref {ext}(b) shows the 3-phase differential currents for LTC = full, switching time = 15.00138s, and switching of 3 Legs of the capacitor bank.
\begin{table}[ht]
\centering
\caption{Parameters for Capacitor Switching\label{tab143}}
{\begin{tabular}{ll}\toprule
Variables  & Values \\ \midrule
Capacitor bank rating & 500,1000,1500 MVAr (3)                  \\
Switching time  & 15s to 15.0153s in steps of 1.38ms (12)    \\
Phase shift & Forward and backward (2)\\
LTC  & 0.2 to full tap in steps of 0.2 (5)\\
& Total = $3\times12\times2\times5$ = 360\\
\hline
\end{tabular}}{}
\end{table}
\subsection{Ferroresonance}  Initiated by faults and switching operations, ferroresonance causes harmonics and overvoltages and may lead to mal-operation of protective relays and damage of power equipment \cite{ferro2}. Mal-operation of the differential relay occurs because of the higher magnitude of current in the HV side than the LV side \cite{ferromanitova}. Besides, the low loss, amorphous core transformer increases the intensity and occurrence of ferroresonance \cite{FERRO3}. Several configurations may lead to ferroresonance in electrical systems. In this paper, one such arrangement has been modeled when one of the phases of a 3-phase transformer is switched off. The parameters and their values for ferroresonance conditions are presented in Table \ref{tab123}.
Fig.\ref {ext}(c) shows the 3-phase differential currents for switching time = 15s and grading capacitance = 0.2$\mu$F simulated between bus2 and bus4.

\begin{table}[ht]
\centering
\caption{Parameters for Ferroresonance\label{tab123}}
{\begin{tabular}{ll}\toprule
Variables  & Values \\ \midrule
Grading capacitance & 0.02$\mu$F to 0.2$\mu$F in steps of 0.02$\mu$F (10)\\
Location & a,b,c phases (3)\\
Switching time         & 15s to 15.016s in steps of 0.69ms   (24) \\
&{Total=$10\times3\times24$=720}\\
\hline
\end{tabular}}{}
\end{table}

\section{Detection, discrimination \& classification Algorithm}\label{sec3}
\subsection{Change detection filter (CDF) for transient detection}
The change in the differential currents in case of transients is detected by a change detection filter (CDF) which calculates the difference between the cumulative sum of modulus of two consecutive cycles.
\begin{equation} CDF (t) = \sum_{x=n_c+t}^{2n_c+t}|Id(x)|-\sum_{x=n_c+t}^{2n_c+t}|Id(x-n_c)|
\end{equation}
where x {equals} sample number which begins at the second cycle, $n_c$ {equals} number of samples in a cycle, ${\{t\}}_{t=1}^{n-{n_c}}$, n {equals} total number of samples, and $Id$ {represents} a, b, and c phase differential currents.

The change detection filter starts logging the data from the instant CDF(t) is greater than a threshold, $th$ in any one of the 3-phases. In normal conditions when there is no transient, the values of CDF(t) are nearer to zero \cite{Dharmapandit2017}. 

\subsection{Feature Extraction $\&$ Selection }\label{feature selection}
Time series analysis of the differential currents helps in the classification and characterization of power system events. Features extracted from these time series are used as input to the machine learning algorithms. Informative and distinctive features that help to classify the events may range from simple statistical functions to complex ones. Researchers have used time-frequency representations like Wavelet Transform \cite{SVM1,SVM2,ann1,dtwt,PQ2,PQ3,PQ6} and Stockwell Transform \cite{PQ1,PQ4,PQ5,PQ6}  to extract features from the non-stationary transients to discriminate inrush and internal faults and for classification of PQ disturbances. In this paper, to differentiate the faults from the other transient disturbances, three time-domain features and  two frequency-domain features have been used.

A comprehensive number of features (794) from different domains are extracted from the 3-phase differential currents obtained from the current transformers, CT1 and CT2 located near bus4 and bus5. The complete list of the features extracted can be found in \cite{tsfresh}. Out of these 794 features, Random Forest is used to rank and select the features with maximum Information Gain to distinguish between the different classes. The most relevant and common features for each of the classification tasks obtained after performing feature ranking belong to the set \textbf{F} = \{F1, F2, F3, F4, F5\}  where, F1 {is} average change quantile, F2 {is} Fourier transform (FT) coefficients, F3 {is} aggregate linear trend, F4 {is} spectral welch density, and  F5 {is} autoregressive coefficients. Only those features of set \textbf{F} which are present in each of the 3-phase differential currents
are used for training the classifiers to detect the faults, localize the faulty units, identify the fault type, and identify the disturbance type (Table \ref{featuretab}). The feature set \textbf{F} is detailed in what follows.

F1, average change quantile calculates the average of absolute values of consecutive changes of the time series inside two constant values $qh$ and $ql$ as
\begin{equation} avg.\  change\ quantile =\frac{1}{n'}\cdot{\sum_{t=1}^{n'-1} |Id_{t+1} - Id_{t}| }
\end{equation} where, $n'$  {equals} number of sample points in the differential current between $qh$ and $ql$, $Id$ {is} a, b, and c phase differential currents with n sample points. 

F2, FT coefficients, (X$|$k) returns the fourier coefficients of 1-D discrete Fourier Transform for real input using  fast FT as
\begin{equation} (X|k)= {\sum_{t=0}^{n-1}Id_t\cdot e(-  \frac{j2 \pi kt}{n} ), k\in Z}
\end{equation}

F3, aggregate linear trend calculates the linear least-squares regression for values of the time series over windows and returns aggregated values of either intercept or standard error. 

F4, spectral welch density uses Welch’s method to compute an estimate of the power spectral density by partitioning the time series into segments and then averaging the periodgrams of the discrete Fourier transform of each segment \cite{welch}.

F5, autoregressive coefficients are the least-square estimates of $\varphi_{i's}$ which are obtained by minimizing Eq.\ref{ar} with respect to $\varphi_0, \varphi_1..., \varphi_P$ and lag P. \begin{equation}\label{ar} \sum_{t=p+1}^{n} [Id_t- \varphi_0 - \varphi_1\cdot Id_{t-1} - ...-\varphi_P \cdot Id_{t-P}]^2 \end{equation}

More than one feature can be extracted from the above time and frequency domain functions by varying their parameters. e.g ($qh,ql$) = (0.8,0.4) \& (0.8,0.2) yields 2 features from change quantile and window length = 5, 10, and 15 would return 3 features of  linear trend.
\subsection{Classifiers}
Tree-based learning algorithms like decision trees, random forest, and gradient boosting are considered among the best and predominantly used supervised learning methods in problems related to data science. These estimators have higher accuracy, stability and are easy to interpret. They can also handle non-linear relationships quite well. DT, RFC, GBC, and SVM has been used to detect and classify the transients.
\subsubsection{Decision Tree}
Decision trees are distribution-free white box Machine Learning models that learn simple decision rules inferred from the feature values. In 1984 Breiman et al. introduced Classification and Regression Trees (CART) \cite{Breiman}. Here, the CART algorithm implemented in scikit-learn is used which constructs binary trees by splitting the training set recursively till it reaches the maximum depth or a splitting doesn't reduce the impurity measure. The candidate parent data $D_p$ is split into $D_l$ and $D_r$ at each node using a feature ($f$) and threshold that yields the largest Information Gain. The objective function IG which is optimized at each split is defined as
\begin{equation}
IG(D_p,f)=I(D_p)-\frac{N_l}{N_p}\cdot I(D_l)-\frac{N_r}{N_p}\cdot I(D_r)
\end{equation}
where, I is impurity measure, ${N_p},{N_l}$ and ${N_r}$ are the number of samples at the parent and child nodes \cite{Quinlan}. Gini, and entropy impurity measures are used in the hyperparameter search.

\subsubsection{Random Forest}
RFC belongs to the family of ensemble trees which builds numerous base estimators and averages their predictions which produces a better estimator with reduced variance. Each tree constitutes a random sample (drawn with replacement) of the training set and the best split is found at each node by considering a subset of input features. The individual trees tend to overfit but averaging the predictions of all trees reduces the variance \cite{Breiman2001}. The main hyperparameters in RFC are no\_of\_estimators (number of trees in the forest), max\_depth (tree depth), and max\_features (feature size to consider when splitting a node). The no\_of\_jobs parameter was also used to parallelize the construction of tree and computation of predictions by using more processing units. Random Forest has also been used during feature selection and ranking  (\ref{feature selection}) to get the relative importance of the features which is measured by the fraction of samples a feature contributes to and the mean decrease in impurity from splitting the samples \cite{phdthesis}.

\subsubsection{Gradient Boosting Classifier}
GBC belongs to the class of ensemble trees which builds the base estimators from weak learners ($w_{p}(x)$) sequentially in a greedy manner which results in a strong estimator \cite{friedman2001} \cite{Mason}. The newly added $w_{p}$ tries to minimize the loss function given $f_{p-1}$, step length ($\lambda_p$), and input ${(x_i,y_i)}_ {i = 1}^  n$.
\begin{equation} \label{eq3}
\begin{split}
f_p(x)=f_{p-1}(x)+ \lambda_p w_p(x)\\
w_p = arg\ \stackunder{min}{w} \sum_{i=1}^{n}L(y_i,f_{p-1}(x_i) +w(x_i))
\end{split}
\end{equation}
The minimization problem is solved by taking the negative gradient of the negative multinomial log-likelihood loss function, L for mutually exclusive classes.
\begin{equation}
f_p(x)=f_{p-1}(x)-\lambda_p\sum_{i=1}^{n}\nabla_{f} L(y_i,f_{p-1}(x_i))
\end{equation} GBC uses shrinkage which scales the contribution of the weak learners by the learning rate and sub-sampling of the training data (stochastic gradient boosting) for regularization. The important hyperparameters of the different GBC classifiers used are the results of grid search on no\_of\_estimators = [5000, 7000, 10000, 12000, 15000], max\_depth = [3,5,7,10,15], and learning\_rate = [0.01, 0.05, 0.07, 0.1]. 
\subsection{Proposed scheme}
\begin{figure}[htb]
\centerline{\includegraphics[width=3.55in, height= 3.3 in]{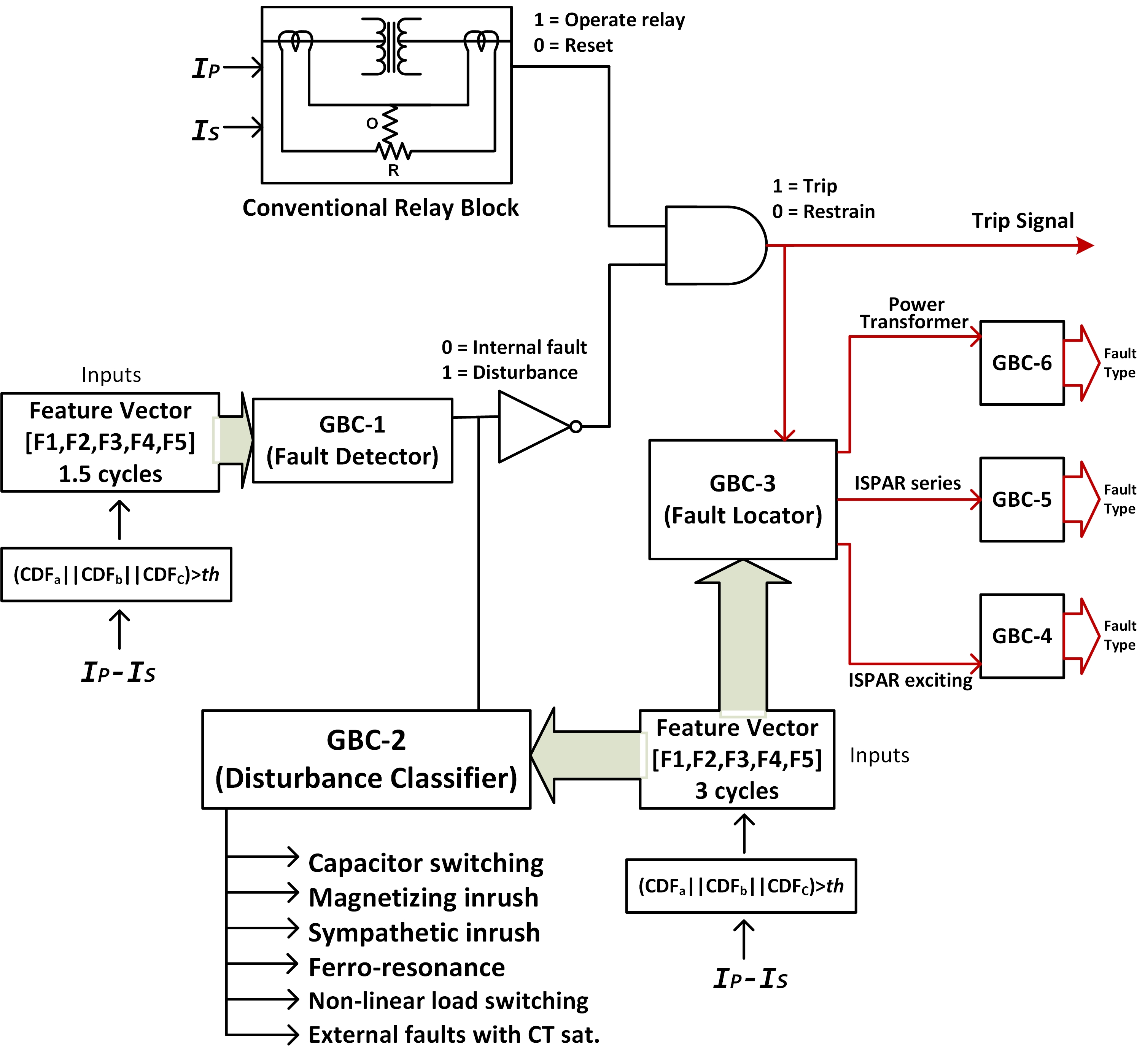}}
\vspace{0mm}\caption{Proposed transient detection and classification algorithm }
\label{flowchart}
\end{figure}
The block diagram description of the CDF and GBC based proposed internal fault detection, fault localization, and transient disturbance classification algorithm is shown in Fig.\ref{flowchart}.
The change detector discovers the change in the 3-phase differential currents ($I_{P}$-$I_{S
}$) if the CDF index in any phase is greater than the threshold, $th$ = 0.05. 1/2 cycle pre-transient and 1 cycle post-transient differential current samples are used to detect an internal fault and 3 post-transient cycles are used for localization of faults and classification of transient disturbances. The scheme consists of a four-level classifier design. The level-1 classifier (GBC-1) consists of the fault detector, which can apply supervisory control over the operation of the conventional differential relay. GBC-1 identifies an internal fault with \say{0} and other transient disturbances with \say{1}. Hence, it governs the working of the trip/restrain function by blocking all other power system transients but an internal fault. The level-2 classifier (GBC-2) does further analysis of the power system events in case the output of GBC-1 is \say{1}.
The GBC-2 can identify the transient disturbance responsible for the mal-operation of the conventional differential relay
(GBC-1 is \say{1} \& Operate relay is \say{1}). The level-3 classifier (GBC-3) locates the faulty transformer unit (PT, ISPAR series, and ISPAR exciting) if the output of GBC-1 is \say{0}. The level-4 classifiers (GBC-4, GBC-5, and GBC-6) further identifies the internal faults in the ISPAR exciting, the ISPAR series and the PT.

\section{Results and Discussion}\label{sec4}
1.5 cycles of 3-phase differential currents are used for detection, and 3 cycles are used for localization and identification of transients from the time of their inception. Thus, at a sampling rate of 10 kHz, 167 data samples per cycle are analyzed. Several factors influence the classification accuracy of an algorithm. Cross-validation and grid search helps in using the data effectively and training the classifier with the best combination of hyperparameters. The data is split randomly into training and test set in a 4:1 ratio. To avoid the problem of overfitting and underfitting of the estimator on the test set, cross-validation is applied on the training data and the hyperparameters are optimized using grid search over a parameter grid. Grid search comprehensively searches for the parameters over the subset of the hyperparameter space of the estimator. The performance of the selected hyperparameters is then tested on the unseen test data that is not used during the training process. Ten-fold stratified cross-validation (rearrangement of the training data in ten folds such that each fold represents every class well) is used to select the model as it is better both in terms of bias and variance \cite{Kohavi}. 
\begin{table}[ht]
\centering
\caption{Internal fault detection with CDF \& GBC-1} \label{tabcdf}
{\begin{tabular}{lllll}\toprule
Fault/Disturbances  & Total & TP  & FN & FP \\
\midrule
Internal Faults   &2107 & 2105 & 2 & 0\\
Disturbances  &1852 & 1852 & 0 & 2\\
\hline
\end{tabular}}{}
\end{table}
\begin{table}[ht]
\centering
\caption{Comparison of performances with and without CDF}\label{tabcompare}
\begin{subtable}[b]{.53\linewidth}
\centering
\caption{Internal fault detection with CDF\label{tabcdf2}}
{\begin{tabular}{ll}\toprule
Classifier & $\bar{\eta}$ \\
\midrule
GBC-1 & \textbf{99.95 \%}\\
DT  & 99.5\%\\
SVM & 99.7\%\\
RFC & 99.9\%\\
\hline
\end{tabular}}
\end{subtable}{}\hfill
\begin{subtable}[b]{.45\linewidth}
\centering
\caption{Internal fault detection without CDF\label{tabwithoutcdf}}
{\begin{tabular}{ll}\toprule
Classifier & $\bar{\eta}$ \\
\midrule
GBC &  \textbf{98.5\%}\\
DT  & 95.3\%\\
SVM & 89.2\%\\
RFC & 94.6\%\\
\hline
\end{tabular}}{}
\end{subtable}{}
\end{table}
\subsection{Internal fault detection}
The detection of internal faults is performed using GBC in two ways, one with the CDF and the other without it. Most authors haven't considered using some technique to detect the change in differential currents in case a transient occurs. Rather they fixed the time of occurrence of the transient events and used this specified inception time to store the disturbance and fault data. However, faults and disturbances are highly unpredictable in time. In this paper, both methods, one considering a specified time (without the use of CDF) and the other with CDF are used to register the data after the inception of transients. The CDF detects the change and registers 1/2 cycle of pre-transient and 1 cycle of post-transient samples. This 1.5 cycle (250 samples) is used to extract the relevant features which are then fed to GBC, SVM, DT, and RFC classifiers. Accuracy is used as the typical metrics to measure the performance of the classifiers. But, it is biased to data imbalance. Since, the classes are imbalanced, balanced accuracy which is defined as mean of the accuracies obtained on all classes and computed as $\bar{\eta}$ = $ \frac{1}{2}\cdot[\frac{TP}{(TP+FN)} + \frac{TN}{(TN+ FP)}]$ for binary classes is used to compute the performance measure where, TP {represents} true positive, TN {represents} true negative, FP {represents} false positive, and FN {represents} false negative \cite{imbalance}.

The performance of the fault detection scheme composed of the GBC-1 and CDF is shown in Table \ref{tabcdf}. $\bar{\eta}$ of 99.95\% is obtained on a training data of 15,835, testing data of 3959, and hyperparameters: learning\_rate = 0.1, max\_depth = 5, and no\_of\_estimators = 7000. The performance of the four classifiers with CDF is shown in Table \ref{tabcdf2}. One cycle of post-fault data is used for training the classifiers for fault detection without the CDF. $\bar{\eta}$ of 98.52\% is obtained with GBC for max\_depth = 7,  no\_of\_estimators = 5000, and learning rate = 0.07. The balanced scores of the four classifiers trained on 80,870 cases and tested on 20,218 cases are shown in Table \ref{tabwithoutcdf}. 18 features from the 3-phase differential currents (Table \ref{featuretab}) 
are used as the input to the classifiers for training the fault detection models with and without CDF. GBC with CDF performed better than without CDF (Table \ref{tabcompare}) as the CDF filtered out the cases where there is no appreciable change in differential currents although a transient event occurred. It is noticed that the CDF could detect the change in differential currents in all internal fault cases except turn-to-turn faults with Rf = 10$\Omega$, LTC = 0.2, and percentage of winding shorted = 20\%. Also, it detected the change for all transient disturbances except sympathetic inrush cases for switching angles from 120$^{\circ}$ to 330$^{\circ}$. On exploring the data it is observed that there is almost no change in the differential currents for these instances. The $w$-g faults for LTC = 0.2, and percentage of winding shorted = 20\% which needs higher sensitivity were detected. It proves the dependability of the scheme for ground faults near neutral of wye grounded transformers (PT and ISPAR exciting) which is again a challenge for conventional differential relays \cite{pstguide}. 

\begin{table}[ht]
\centering
\caption{Localization of faulty transformer unit}
\setlength{\tabcolsep}{3.3pt}
\begin{subtable}[l]{.55\linewidth}
\centering
\vspace{-5pt}
\caption{ Localization with GBC-3\label{tabfaulty_trans1}}
{\begin{tabular}{lllll}\toprule
  Transformer  &   Total &   TP  &   FN &   FP\\
\midrule
  ISPAR Exciting&  2937 &   2899 & 38 &  8\\
  ISPAR Series & 7402 & 7383 &  19 &  17\\
 {PT} & 7287 & 7287 & 0&  32\\
\hline
\end{tabular}}{}
\end{subtable} \hfill
\begin{subtable}[r]{.37\linewidth}
\centering
\vspace{4pt}
\caption{Comparison of performances\label{tabfaulty_trans2}}
{\begin{tabular}{ll}\toprule
  Classifier & $\bar{\eta}$ \\
\midrule
  GBC-3 & \textbf{99.5\%}\\
  DT  &98.6\%\\
  SVM & 88.9\%\\
  RFC &98.7\%\\ \hline
\end{tabular}}
\end{subtable}
\end{table}

\subsection{Identification of faulty unit $\&$ internal fault type}

Once it is confirmed that an internal fault has been detected, the locations of those internal faults are determined. 3 cycles of post-fault differential current samples are used to locate the faulty transformer unit (PT or ISPAR Exciting or ISPAR Series) and determine the type of fault. GBC, SVM, DT, and RFC are used to identify the faulty unit and further locate and pinpoint the type of fault in the PT and ISPAR units. 
$\bar\eta$ and accuracy computed as $\eta$ = $\frac{(TP+TN)}{(TP+FN+TN+FP)}$, are used as the metrics to measure the performance of the estimators for localization of faulty unit and identification of internal fault type, respectively.

\begin{table}[ht]
\centering
\caption{Comparison of identification performances of internal fault type}
\begin{subtable}[b]{.3\linewidth}
\centering
\caption{Exciting unit \label{tabexc}}
{\begin{tabular}{ll}\toprule
Classifier & $\eta$ \\
\midrule
GBC-4 & \textbf{99.2\%}\\
DT  & 98.6\%\\
SVM & 94.8\%\\
RFC & 98.9\%\\
\hline
\end{tabular}}{}
\end{subtable}{}\hfill
\begin{subtable}[b]{.3\linewidth}
\centering
\caption{Series unit\label{tabseries}}
{\begin{tabular}{ll}\toprule
Classifier & $\eta$ \\
\midrule
GBC-5 &  \textbf {98.0\%}\\
DT  &94.7\%\\
SVM & 90.7\%\\
RFC &96.9\%\\
\hline
\end{tabular}}
\end{subtable}{}\hfill
\begin{subtable}[b]{.3\linewidth}
\centering
\caption{PT\label{tabpt}}
{\begin{tabular}{ll}\toprule
Classifier & $\eta$ \\
\midrule
GBC-6 & \textbf{99.2\%}\\
DT  & 98.9\%\\
SVM & 94.0\%\\
RFC & 97.8\%\\
\hline
\end{tabular}}{}
\end{subtable}{}
\end{table}

\subsubsection{Localization of faulty unit}
To locate the faulty transformer unit 70,502 fault cases are trained and 17,626 cases are tested. 18 features are used to train the classifiers (Table \ref{featuretab}).
GBC-3 with hyperparameters: no\_of\_estimators = 5000, learning\_rate = 0.07, and max\_depth = 10 gives $\bar\eta$ of 99.48\%. Table \ref{tabfaulty_trans1} shows the localization results using GBC-3 and Table \ref{tabfaulty_trans2} compares the $\bar\eta$ of the four different classifiers.

\subsubsection{Identification of internal fault type}
The internal faults in the ISPAR series, ISPAR exciting and the PT are further classified into $w_{a}$-g, $w_{b}$-g, $w_{c}$-g, $w_{a}$-$w_{b}$-g, $w_{a}$-$w_{c}$-g, $w_{b}$-$w_{c}$-g, $w_{a}$-$w_{b}$, $w_{a}$-$w_{c}$, $w_{b}$-$w_{c}$, turn-to-turn, winding-to-winding, and very rare $w_{a}$-$w_{b}$-$w_{c}$ and $w_{a}$-$w_{b}$-$w_{c}$-g faults. 21 features from 3 cycles of the 3-phase differential currents are used as the input to the estimators (Table \ref{featuretab}).
 Tables \ref{tabexc}, \ref{tabseries}, and \ref{tabpt} compare the performances of GBC, RFC, DT, and SVM classifiers for ISPAR exciting, ISPAR series, and the PT respectively.

To identify the internal faults in ISPAR exciting 14,688 fault cases are used to train and test the four classifiers. GBC-4 trained with hyperparameters of max\_depth = 5, no\_of\_estimators = 7000, and learning\_rate = 0.1 achieved the best accuracy of 99.18\%. 
For the identification of internal faults in ISPAR series 36,720 cases are used to train and test the classifiers. GBC-5 trained with learning\_rate = 0.05, max\_depth = 7, and no\_of\_estimators = 5000 gives an accuracy of 98.0\%. 
Similarly, for PT the classifiers are trained \& tested on 36,720 fault cases. GBC-6 achieved the best accuracy of 99.2\% obtained by training the hyperparameters on learning\_rate = 0.05, no\_of\_estimators = 5000, and max\_depth = 5.
The identification accuracy obtained in the ISPAR series is lower than in PT and ISPAR exciting because the secondary side of ISPAR series is delta connected. Hence, one type of fault on the primary side confuses with another type on the secondary side. 

\subsection{Identification of disturbance type} 
\begin{figure}[htb]
\centerline{\includegraphics[width=3.54 in, height= 2.3 in]{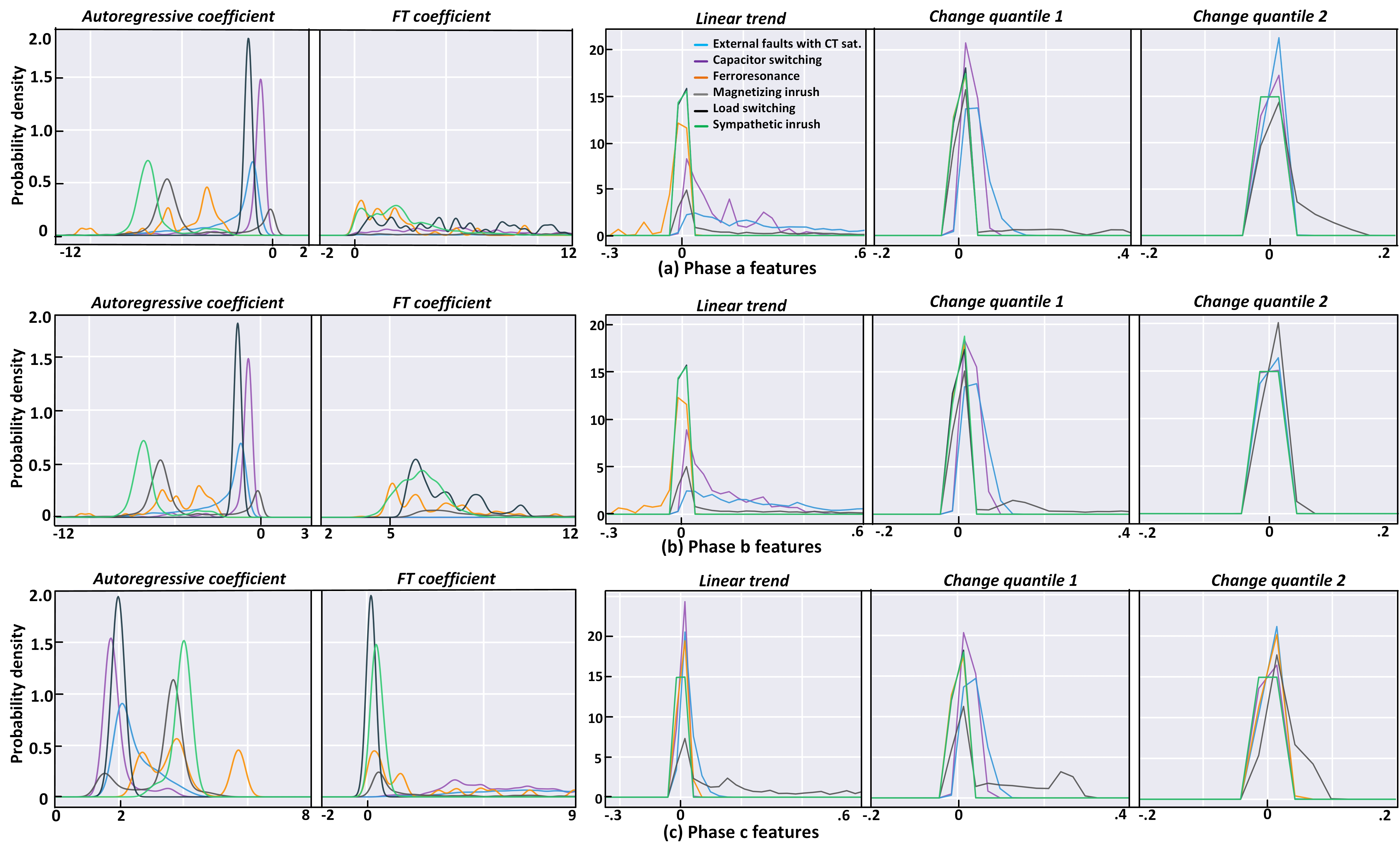}}
\vspace{0mm}\caption{Kernel Density Estimate plots showing the probability distribution of the 5 selected features for the 6 transient disturbances in (a) phase a, (b) phase b, and (c) phase c.}
\label{kdeplot}
\end{figure}

\begin{figure}[htb]
\centerline{\includegraphics[width=3.0 in, height= 1.75 in]{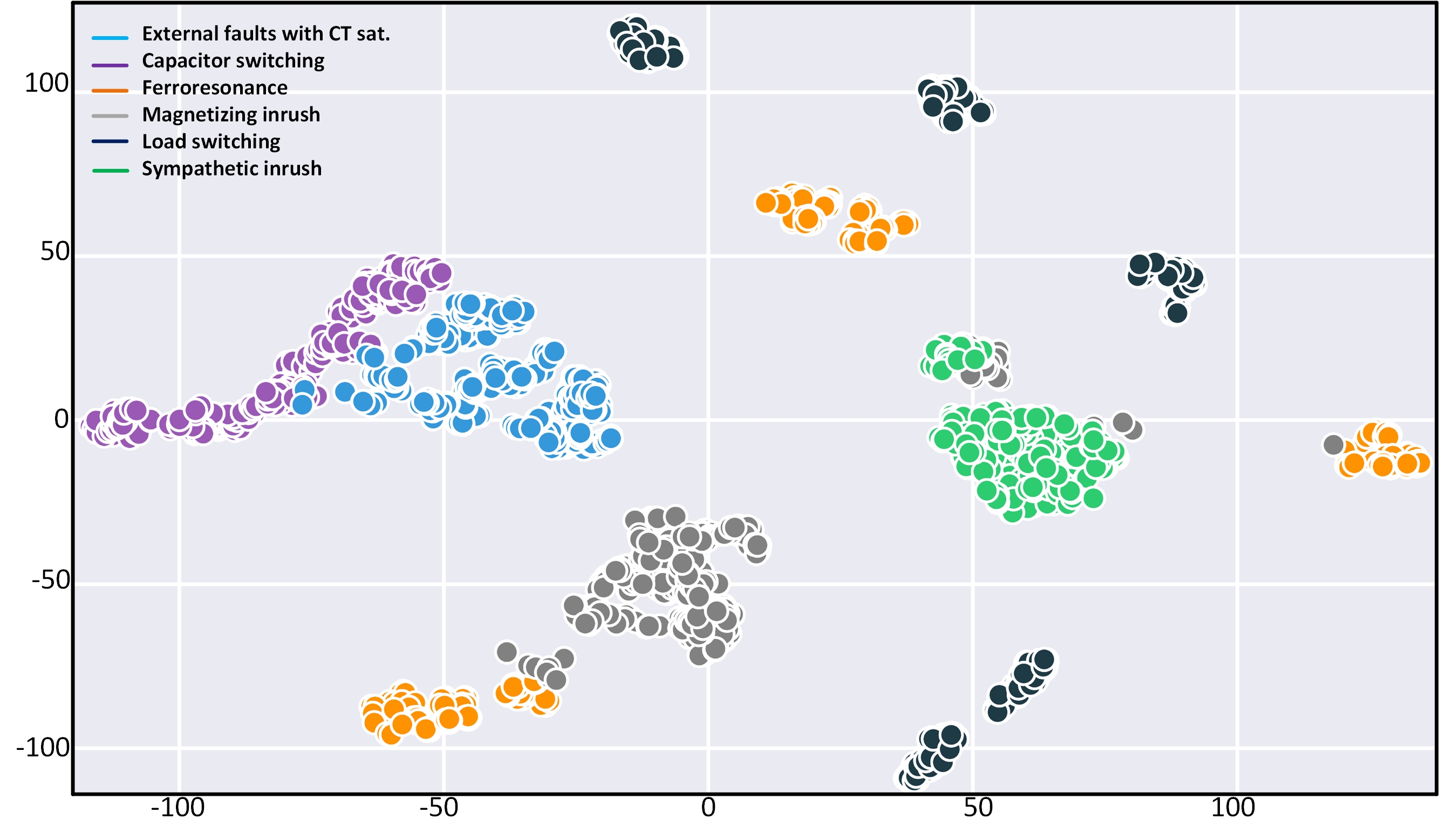}}
\vspace{0mm}\caption{2D scatterplot of the input features for transient disturbances}
\label{scatter}
\end{figure}

The various disturbances: magnetizing inrush, sympathetic inrush, ferroresonance, external faults with CT saturation, capacitor switching, and non-linear load switching are also classified using 3 cycles of post-transient samples after they are differentiated as no-fault by the fault detection scheme. 
15 features are used as input to the classifiers in this case (Table \ref{featuretab}).
{It's always useful to know the probabilities of the input features taking on various real values. Parzen–Rosenblatt window method is used to estimate the underlying probability density of the 5 features for the six different disturbances in phases a, b, and c. Fig.\ref{kdeplot} shows the kernel density estimation plots for the chosen features. Gaussian is used as the kernel function to approximate the univariate features with a bandwidth of 0.2 for autoregressive coefficient, and FT coefficient and a bandwidth of 0.01 for aggregate linear trend, and avg. change quantile 1 and 2. It is observed that probability density functions of autoregressive and FT coefficients are a mixture of multiple normal distributions with varying standard deviation and mean whereas linear trend, and change quantiles are unimodal with means near zero and smaller standard deviations. Table \ref{feature1} and \ref{feature2} shows the values of mean ($\mu$), variance ($\sigma^2 $), skewness ($\Tilde{\mu_3}$), and kurtosis ($\kappa$) of the 5 features for magnetizing inrush and CT saturation during external faults respectively in phases a, b and c. Because of space limitations, the feature statistics of only these two transients are shown. 
Furthermore, to visualize the 15-dimensional input data in a 2-dimensional plane, the T-distributed Stochastic Neighbor Embedding dimensionality reduction technique has been used which preserves much of the significant structure in the high-dimensional data while mapping in the 2-dimension  \cite{tsne}. Fig.\ref{scatter} shows the clusters of similar transients (300 instances each) and also the relationships between different group of transients as a scatter plot.}

The Table \ref{tabdisturbance1} shows the classification results using GBC-2. Table \ref{tabdisturbance2} compares the results of GBC with RFC, DT, and SVM. The classifiers are trained on 10,368 cases and tested on 2592 cases. $\bar{\eta}$ of 99.28\% is obtained with GBC-2 having hyperparameters: no\_of\_estimators = 5000, learning\_rate = 0.7, and max\_depth = 3.

\begin{table}[ht]
\centering
\caption{Low and high-order statistics of the 5 selected features}
\begin{subtable}[b]{\linewidth}
\centering
\caption{Magnetizing inrush\label{feature1}}
\setlength{\tabcolsep}{2.1pt}
\tiny
\begin{tabular}{@{}lllllllllllllllllllll@{}}
\toprule
\multirow{2}{*}{} & \multicolumn{4}{c}{\begin{tabular}[c]{@{}c@{}}autoregressive\\ coefficient\end{tabular}} & \multicolumn{4}{c}{\begin{tabular}[c]{@{}c@{}}FT\\ coefficient\end{tabular}} & \multicolumn{4}{c}{\begin{tabular}[c]{@{}c@{}}linear \\ trend\end{tabular}} & \multicolumn{4}{c}{\begin{tabular}[c]{@{}c@{}}change\\ quantile 1\end{tabular}} & \multicolumn{4}{c}{\begin{tabular}[c]{@{}c@{}}change \\ quantile 2\end{tabular}} \\ \cmidrule(l){1-21} 
                  & $\mu  $ & $\sigma^2$  &      $\Tilde{\mu_3}  $    & $\kappa$         &  $\mu  $ & $\sigma^2$  &      $\Tilde{\mu_3}  $    & $\kappa$&$\mu  $ & $\sigma^2$  &      $\Tilde{\mu_3}  $    & $\kappa$                  &  $\mu  $ & $\sigma^2$  &      $\Tilde{\mu_3}  $    & $\kappa$                 &   $\mu  $ & $\sigma^2$  &      $\Tilde{\mu_3}  $    & $\kappa$                                        \\ \midrule
\it {ph a}                  &       -4.6                 &        5.9               &     .74               &        -.35            &       1e3            &       2e6            &       1.3            &        .74          &        2.6           &    14               &      1.9           &     3.9             &      .02            &        9e-4            &        1.9           &        2.9           &       .04             &    .008                &  2.5                  &      5.1             \\
\it {ph b}                 &        -4.7                &    6.2                  &.65                    &   -.13                &      502             &     3e5              &      1.0             &       -.06           &       2.6            &    13               &      1.8            &         3.2         &         .01           &        8e-5            &        1.5           &         1.06          &     .04               &         .01           &          2.5          &           5.7        \\

\it {ph c}                &      3.3                  &  .72                     &      -1.1              &      .33              &       33            &        1e3          &     1.1              &        .25          &      .19             &    .05               &      1.0           &    -.35              &        .02            &       5e-4             &    .90               &     -.45              &        .08            &   .009                 &         .88           &     -.84              \\ \bottomrule
\end{tabular}
\end{subtable}
\begin{subtable}[b]{\linewidth}
\vspace{2 mm}
\centering
\caption{ CT saturation during external faults\label{feature2}}
\setlength{\tabcolsep}{2.1pt}
\tiny
\begin{tabular}{@{}lllllllllllllllllllll@{}}
\toprule
\multirow{2}{*}{} & \multicolumn{4}{c}{\begin{tabular}[c]{@{}c@{}}autoregressive\\ coefficient\end{tabular}} & \multicolumn{4}{c}{\begin{tabular}[c]{@{}c@{}}FT\\ coefficient\end{tabular}} & \multicolumn{4}{c}{\begin{tabular}[c]{@{}c@{}}linear \\ trend\end{tabular}} & \multicolumn{4}{c}{\begin{tabular}[c]{@{}c@{}}change\\ quantile 1\end{tabular}} & \multicolumn{4}{c}{\begin{tabular}[c]{@{}c@{}}change \\ quantile 2\end{tabular}} \\ \cmidrule(l){1-21} 
                  & $\mu  $ & $\sigma^2$  &      $\Tilde{\mu_3}  $    & $\kappa$         &  $\mu  $ & $\sigma^2$  &      $\Tilde{\mu_3}  $    & $\kappa$&$\mu  $ & $\sigma^2$  &      $\Tilde{\mu_3}  $    & $\kappa$                  &  $\mu  $ & $\sigma^2$  &      $\Tilde{\mu_3}  $    & $\kappa$                 &   $\mu  $ & $\sigma^2$  &      $\Tilde{\mu_3}  $    & $\kappa$                                        \\ \midrule
\it {ph a}                  &       -2.3                 &        3.1               &     -1.6               &        1.97            &      190            &       4e4            &       1.6            &        2.5         &        .47           &    .23              &      1.64          &     2.9             &      .04            &        6e-4            &        1           &        .98           &       .003             &    9e-6                &  1.8                  &     3.6           \\
\it {ph b}                 &        -2.2               &    3                 &-1.6                   &   1.95               &      261             &     2e4              &      .81            &      .35          &     .47 & .24&1.55&2.3&7e-4           &    9e-7               &     3.5            &        19      &         .04          &        5e-4            &        .6          &       -.42            \\

\it {ph c}                &     2.5                  &  .4                     &      1.1            &      .27              &       16           &        150         &     1.7             &        2.9          &      .02            &       5e-4           &    1.5 & 2.4            &        .003           &       8e-6             &    1.7               &     4.7              &        .04           &   4e-4                 &         .67           &     -.14              \\ \bottomrule
\end{tabular}
\end{subtable}
\end{table}

\begin{table}[ht]
\centering
\caption{Identification of transient disturbances with GBC-2\label{tabdisturbance1}}
{\begin{tabular}{lllll}\toprule
Disturbances   & Total & TP  & FN &   FP\\
\midrule
Magnetizing inrush &  365& 357 & 8 &0\\
Sympathetic inrush  &336 & 336 &0 &8\\
Capacitor switching  & 73 & 72 &1 & 1\\
Ferroresonance & 133 & 132 & 1 & 0\\
Load switching & 69 & 69 & 0 & 0\\
External faults  & 1616 & 1615 & 1 & 2\\
\hline
\end{tabular}}{}
\end{table}

\begin{table}[ht]
\centering
\noindent\begin{minipage}{1.3in}\centering
\caption{Performance comparison of identification  of transient disturbances}\label{tabdisturbance2}
\begin{tabular}{lc}\toprule
Classifier & $\bar{\eta}$ \\
\midrule
GBC-2 & \textbf{99.28\%}\\
DT  & 98.09\%\\
SVM & 98.23\%\\
RFC & 98.89\%\\ 
\hline
\end{tabular}
\end{minipage}\hspace{0.8cm}
\begin{minipage}{1.4in}\centering\setlength{\tabcolsep}{2.1 pt}
\caption{Misclassification between ISPAR \& PT}\label{tabptpst}
\begin{tabular}{lcccc}\toprule
Faults   & Total & TP  & FN & FP \\\midrule
PT   &7262 & 7262 & 0 & 13\\
ISPAR  &10364 & 10351  & 13 & 0\\\hline
\end{tabular}
\end{minipage}
\end{table}

\subsection{Discriminate faults in PT \& ISPAR}
The PAR controls the power flow through a line and when connected with a PT, it reduces the magnitude of the differential currents and their harmonic contents and alters the wave shapes due to the additional phase shift for the PT in case of external faults. With internal faults, such changes in the differential currents are lesser. 70,502 internal fault cases are trained and 17,626 cases are tested on 18 features to verify how effectively the GBC discriminates the internal faults in the PT and the ISPAR.
The table \ref{tabptpst} shows the classification errors with learning\_rate of 0.05, max\_depth = 9, and no\_of\_estimators = 5000. The balanced accuracy of 99.9\% shows that the GBC is capable of distinguishing these faults even in an interconnected network.

\subsection{Performances on balanced and imbalanced data}
Machine learning algorithms are more reliable when they operate on a balanced dataset. To adjust the data distribution of classes and remove class imbalance, under-sampling of majority classes and over-sampling of minority classes is performed. Synthetic Minority Over-Sampling Technique (SMOTE) \cite{smote} is used to create minority synthetic data considering k-nearest neighbours and NearMiss algorithm is used for under-sampling the majority classes avoiding information loss. The table \ref{featuretab} shows the balanced accuracy/ accuracy for detecting the internal faults, identifying the faulty units and type of faults in those units, and identifying the disturbances with and without using SMOTE and NearMiss. It is observed that the accuracies obtained with SMOTE and NearMiss algorithm for the different classification tasks are similar to those obtained by training the GBCs without them. Table \ref{featuretab} also gives the information about the time and frequency domain features $({\{F_i\}}_{i=1}^{5})$ that has been used to train the different GBC classifiers for the different classification tasks. 

\begin{table}[h!]
\centering
\caption{ Input features and performance of different GBC classifiers with and without SMOTE analysis\label{featuretab}}
\setlength{\tabcolsep}{3.3pt}
\begin{tabular}{lcccccccc}\toprule
 Classification task  &F1&F2&F3&F4&F5& $\sum Fi$$\times3$& \multicolumn{1}{c}{\begin{tabular}[c]{@{}c@{}}$\bar{\eta}$\textbackslash${\eta}$ \\  (\%)\end{tabular}}   &  \multicolumn{1}{c}{\begin{tabular}[c]{@{}c@{}}\textit{$\bar{\eta}$ using}\\ \textit{SMOTE} \end{tabular}}  \\
 \midrule
detect  faults  &2&1&1&1&1&18& 99.9&99.9\\ 
locate faulty units &2  &2&2&-&-&18& 99.5&99.6\\ 
 identify faults (series)  &3&1&2&1&-&21& 98.0&98.2\\ 
 identify faults(exciting)  &3&2&2&-&-&21& 99.2&99.1\\ 
 identify faults (PT)  &3&2&2&-&-&21&99.2&99.1\\ 
 identify transients   &2&1&1&-&1&15& 99.3&99.4\\ \bottomrule
\end{tabular}{}
\end{table}

\subsection{Effect of different rating and transformer connections}
It is not necessary to train the fault detection scheme for different rating and connection of the PTs, rating of ISPAR, and variation in other parameters. In order to validate the effectiveness of the proposed scheme with variation in different system parameters, new internal faults and other transient cases are simulated again  with 400 MVA, Y$\Delta$ connected PTs and 400 MVA ISPARs. The fault resistance, LTC, fault type, and fault inception time are altered to generate the internal fault cases and switching time, firing angle, LTC, etc. are altered to generate the transient cases to test the same GBC-1 model trained using 500 MVA, YY connected PTs and 500 MVA ISPARs. It is observed from Table \ref{dif connection} that the proposed scheme gives a balanced accuracy of 99.3\% which is compatible with the accuracy obtained when trained and tested at 500 MVA and YY connection.

\begin{table}[]
\caption{Performance for 400-MVA \& Y$\Delta$ connection\label{dif connection}}
\setlength{\tabcolsep}{2.2pt}
\begin{tabular}{@{}clcccc@{}}
\toprule
\begin{tabular}[c]{@{}c@{}}Fault/\\ Disturbances\end{tabular}                           & \multicolumn{1}{c}{\begin{tabular}[c]{@{}c@{}}Faults/\\ Abnormalities\end{tabular}} & Total & TP   & FN & \multicolumn{1}{c}{\begin{tabular}[c]{@{}c@{}}$\bar{\eta}$\\ (\%)\end{tabular}} \\ \midrule
\multirow{4}{*}{\begin{tabular}[c]{@{}c@{}}Internal \\ faults\\ (3072)\end{tabular}} & (a) ph \& g, T-T, W-W faults (PT)                                                         & 1200  & 1200 & 0  &      100               \\
                                                                                        & (b) ph \& g, T-T, W-W faults (Series)                                                     & 1200  & 1200 & 0  &  100                   \\
                                                                                        & (c) ph \& g, T-T, W-W faults (Exciting)                                                   & 672   & 672  & 0  &     100                \\ \cmidrule(l){2-6}
                                                                                        & (d) Total  =  (a)+ (b)+ (c)                                                               & 3072  & 3072 & 0  &    100                 \\ \midrule
\multirow{6}{*}{\begin{tabular}[c]{@{}c@{}}Other \\ disturbances\\ (876)\end{tabular}}  & (e) Capacitive switching                                                                & 60    &   51   & 9   &     85                \\
                                                                                        & (f) External faults with CT saturation                                                & 528   &   525   &  3  &     99.4                \\
                                                                                        & (g) Ferroresonance                                                                      & 24    &  24    &   0 &    100                \\
                                                                                        & (h) Magnetizing inrush                                                                  & 60    & 60     & 0  &      100               \\
                                                                                        & (i) Load switching                                                                      & 144   &  144    &  0  &      100               \\
                                                                                        & (j) Sympathetic inrush     &   60   &  60  &    0 & 100                 \\  \cmidrule(l){2-6}
\multicolumn{1}{l}{}                                                                    & (k) Total = (e)+ (f)+ (g)+ (h)+ (i)+ (j)                                                                               & 876   &   864   &  12  &     98.6                \\ \midrule
Total (3948)                                                                             & Total faults and disturbances =   (d)+(k)                                                               &  3948     &  3936    &  12  &   99.3                  \\ \bottomrule
\end{tabular}
\end{table}

\subsection{Effect of Signal Noise \& CT saturation}
\begin{figure}[htb]
\centerline{\includegraphics[width=3.52 in, height= 0.95 in]{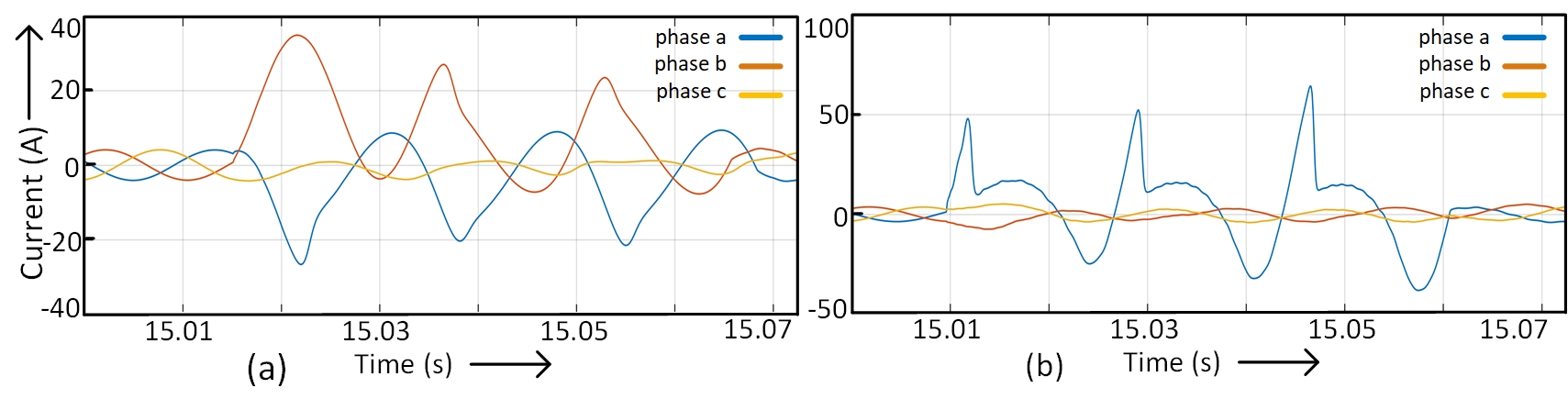}}
\vspace{0mm}\caption{3-phase differential currents with CT saturation for (a) $w_{a}$-$w_{b}$-g fault in transformer primary and (b) $w_{a}$-g fault in series primary}
\label{cts}
\end{figure}

{In order to analyse the effect of noise in the differential currents on the proposed fault detection scheme white Gaussian noise of different levels measured in terms of Signal-to-Noise-ratio (SNR) are added to the training and testing cases for fault detection \cite{PQ2,PQ4,PQ5,PQ6}. Table \ref{noise} shows the accuracy of the GBC for different levels of noise on 5000 cases of internal faults and other disturbances each. It is observed that as the level of noise increases the ${\eta}$ of the classifier dips, but still always above 93.8\% \footnotesize{($\frac{90.4+97.2}{2}$)}. \normalsize The ${\eta}$ changes from 99.4\% to 93.8\% as the SNR is varied from $\infty$ to 10dB. It is also observed from the table that the misclassification of internal faults increases as the SNR is decreased whereas the misclassifications are nearly the same for other disturbances as SNR is decreased from 30dB to 10dB. Moreover, to examine the effect of CT saturation the secondary side impedance (burden and CT secondary impedance) which has the major influence over the level of saturation is changed. $\eta$  of 99.5\% is obtained with GBC on 5000 cases of internal faults and other disturbances each. Fig.\ref{cts} shows the 3-phase differential currents with CT saturation for faults in T1 and ISPAR1.}

\begin{table}[]
\centering
\caption{Effect of Noise\label{noise}}
\setlength{\tabcolsep}{4 pt}
\begin{tabular}{@{}cccccc@{}}
\toprule
\multirow{2}{*}{\begin{tabular}[c]{@{}c@{}}Fault/\\Disturbances\end{tabular}} & \multirow{2}{*}{\begin{tabular}[c]{@{}c@{}}SNR (dB)\end{tabular}} & \multirow{2}{*}{\begin{tabular}[c]{@{}c@{}}Number\\ of cases\end{tabular}} & \multicolumn{2}{c}{Predicted class} & \multirow{2}{*}{\begin{tabular}[c]{@{}c@{}}Accuracy\\ (\%)\end{tabular}} \\ \cmidrule(lr){4-5}
\multicolumn{2}{c}{}                                                                                          &                                                                            & Faults        & Disturbances        &                                                                        \\ \midrule
\multirow{4}{*}{\begin{tabular}[c]{@{}c@{}}Internal \\ faults\end{tabular}}                 & $\infty $             & 1010                                                                       &    1001         & 9                &         99.2                                                               \\
                                & 30             &   1035                                                                         &      1005       &       30               &    97.1                                                                  \\
                                                                                            & 20              &       1008                                                                     &      934         &       74              &  92.7                                                                      \\
                                                                                            & 10              &     984                                                                      &     891        &      93              &    90.4       \\ \midrule                                                              
\multirow{4}{*}{\begin{tabular}[c]{@{}c@{}}Other \\ disturbances\end{tabular}}              &  $\infty $             &   990                                                                     &       3       &         987            &    99.7                                                                    \\
                                 &   30              &     965                                                                      &     24          &       941               &    97.6                                                                   \\
                                                                                            &    20             &  992                                                                         &    26          &    966             &   97.4                                                                     \\
                                                                                            &  10               &  1016                                                                         &   28         &    988               &      97.2         \\ \midrule                                                          
\end{tabular}
\end{table}

\subsection{Execution Time}
The execution time-averaged over 10 runs for the feature extraction, training, and testing of the GBC classifiers for detection of internal faults, identifying the faulty unit and type of fault, and identifying the transient using one CPU core is reported using the in-built library in python (Table \ref{ptime}).
The fault/no-fault decision includes the time to compute the feature and testing a single instance with GBC-1 which adds to 8.7ms with the CDF. Thus, the proposed scheme has a processing time of 25.37ms (16.67+1.7+7) or $\approx 1\frac{1}{2}$ cycle to detect a fault. Considering that these computations can be further optimized for example by converting Python and MATLAB code to a compiled low-level language such as C, the fault detection and localization, and transient identification schemes are suitable for future real-time implementation. {The DT, SVM, RFC, and GBC classifiers are built in Python 3.7 using Scikit-learn framework \cite{scikit2} while the CDF is implemented in MATLAB 2017. The pre-processing of the data is done in Python and MATLAB. All PSCAD simulations are carried out on Intel Core i7-6560U CPU @ 2.20 GHz and 8 GB RAM  and the classifiers are run on Intel Core i7-8700 CPU @ 3.20 GHz and 64 GB RAM.}

\begin{table}[ht!]
\caption{Execution time of the GBC classifiers (seconds)}\label{ptime}
\setlength{\tabcolsep}{1.3 pt}
\begin{tabular}{lcccccc}
\toprule
\multirow{2}{*}{Classification task} & \multirow{2}{*}{\begin{tabular}[c]{@{}c@{}}Training \\ instances\end{tabular}} & \multirow{2}{*}{\begin{tabular}[c]{@{}c@{}}Testing\\ instances\end{tabular}} & \multirow{2}{*}{\begin{tabular}[c]{@{}c@{}}Training\\ time\end{tabular}} & \multicolumn{2}{l}{Testing time} & \multirow{2}{*}{\begin{tabular}[c]{@{}c@{}}Feature\\ extraction \\ time\end{tabular}} \\ \cmidrule(lr){5-6}
                                     &                                                                                &                                                                              &                                                                          & One             & All            &                                                                                       \\ \addlinespace[1mm]

\midrule

 {detect faults  }&  80870 & 20218 & 1506 &   0.0024 & 1.1&  0.007      \\
 detect faults with \scriptsize{CDF}&  15835 & 3959 & 19 &   0.0017 &.061&  0.007       \\
 {locate faulty units }&  70502 & 17626 & 1232  & 0.0049 & 1.2 & 0.0088            \\
 {identify faults (series) }& 29376 & 7344 &  1341  &0.016 & 2.1 &  0.0111            \\
 {identify faults(exciting) }&  11750 & 2938 &  357 &0.013 & 0.48 & 0.0108             \\
 {identify faults (PT)}& 29376 & 7344 & 2712  &0.026 &5.1 & 0.0108           \\

 {identify transients }  & 10368 & 2592 & 72.3  & 0.004 &  0.09   & 0.0065     \\ \midrule
\end{tabular}
\end{table}

 This work distinguishes faults from the six transients ($\bar{\eta}$ = 99.95\%), locates the faulty unit ($\bar{\eta}$ = 99.5\%), identifies the fault type (${\eta}$ $\approx$ 99\%) and six other transients ($\bar{\eta}$ = 99.3\%) for the ISPAR and PT in an interconnected system, whereas the publications \cite{pallav} \cite{pallav2} in the literature focused only on the ISPAR. In \cite{pallav}, only the internal faults in ISPAR were identified and in \cite{pallav2}, the internal faults were differentiated from magnetizing inrush using WT and then the internal faults were identified. In addition to its broader functionality, the current work improves the accuracy from an average of 98.76 [22] and 97.7\%  \cite{pallav2} to 99.2\%.
 
\section{Conclusion}
In this paper, the task of discrimination of internal faults and other transient disturbances in a 5-bus interconnected power system for PTs and PARs is presented. The internal faults including turn-to-turn and winding-to-winding faults in the ISPAR and the PT are distinguished from magnetizing inrush, sympathetic inrush, ferroresonance, external faults with CT saturation, capacitor switching, and non-linear load switching transients. A change detector is used to detect the change in the 3-phase differential currents in case a transient event occurs and registers the current samples for detection and classification purposes. Five most relevant time and frequency domain features, selected from the differential currents on the basis of Information Gain are used to train the DT, RFC, GBC, and SVM classifiers. The fault detection scheme comprising of the CDF and GBC gives an accuracy of 99.95\% on 19,794 transient cases obtained by varying different parameters for the internal faults and other transient disturbances confirming its dependability for internal faults and security against transient disturbances. Once an internal fault is detected and a trip signal is issued using 1.5 cycles, the faulty transformer unit (PT, ISPAR series, or ISPAR exciting unit) and type of internal faults in those units are also identified in 3 cycles. Furthermore, the type of transient disturbance is determined in case the fault detection scheme detects a transient other than internal faults. {The validity of the scheme is also established for different rating and connection of the transformers, CT saturation, and SNR ratio of 30dB to 10dB in the differential currents.}
The proposed fault detection strategy can work together with a conventional differential relay offering supervisory control over its operation and thus avoid false tripping. The transient detection and identification accuracies obtained are among the best even when compared with results from works on isolated and simple networks.

\appendix 
Fortran script for two-winding transformer\label{a1}

\begin{table}[ht]
\footnotesize{}
 \begin{tabular}{ll}\toprule
  1. NW = 4 &         18. L2l = Lk1/2*$fb$\\
  2. $I_m2$ = $I_m1$ = $I_m$ &           19. L3l = Lk2/2*$fc$\\
  3. $fa=fault1*0.01 $ &       20. L4l = Lk2/2*$fd$\\
  4. $fb=1.0-fa$ &       21. L1m = (v1/(w*$I_m1$*i1))*$fa*fa$\\
  5. $fc=fault2*0.01$ &       22. L2m = (v1/(w*$I_m1$*i1))*$fb*fb$\\
  6. $fd=1.0-fc$ &          23. L3m = (v2/(w*$I_m2$*i2))*$fc*fc$\\
  7. i1 = MVA/v1 &        24. L4m = (v2/(w*$I_m2$*i2))*$fd*fd$\\
  8. i2 = MVA/v2 &        25. Lx = L1l + L1m\\
  9. z1 = v1/i1 &           26. Ly = L2l + L2m\\
  10. z2 = v2/i2  &          27. Lz = L3l + L3m\\
 11. w = 2*pi*f   &             28. Lw = L4l + L4m\\
  12. l1 = v1/(w*$I_m1$*i1) &           29. Mxy = sqrt(L1m*L2m)\\
  13. l2 = v2/(w*$I_m2$*i2) &            30. Mxz = sqrt(L1m*L3m)\\
  14. Lk1 = Xl*z1/w &          31. Mxw = sqrt(L1m*L4m)\\
   15. Lk2 = Xl*z2/w  &         32. Myz = sqrt(L2m*L3m)\\
   16. tr = v1/v2 & 33. Myw = sqrt(L2m*L4m)\\
  17. L1l = Lk1/2*$fa$ & 34. Mzw = sqrt(L3m*L4m)\\ \hline
\end{tabular}
\end{table}
\bibliographystyle{IEEEtran}
\bibliography{references}
\end{document}